\documentclass[10pt,letterpaper]{article}
\usepackage[top=0.85in,left=2.75in,footskip=0.75in]{geometry}
\usepackage{changepage}
\usepackage[utf8]{inputenc}
\usepackage{textcomp,marvosym}
\usepackage{fixltx2e}
\usepackage{amsmath,amssymb}
\usepackage{cite}
\usepackage{nameref}
\usepackage[right]{lineno}
\usepackage{microtype}
\DisableLigatures[f]{encoding = *, family = * }
\usepackage{rotating}

\usepackage{bm}
\usepackage{color}

\raggedright
\usepackage[aboveskip=1pt,labelfont=bf,labelsep=period,justification=raggedright,singlelinecheck=off]{caption}
\makeatletter
\renewcommand{\@biblabel}[1]{\quad#1.}
\makeatother

\date{}

\begin{document}
\vspace*{0.35in}

\begin{flushleft}
{\Large
\textbf\newline{SCOTTI: Efficient Reconstruction of Transmission within Outbreaks with the Structured Coalescent}
}
\newline
\\
Nicola De Maio\textsuperscript{1,2,*},
Chieh-Hsi Wu\textsuperscript{2},
Daniel J Wilson\textsuperscript{1,2,3}
\\
\bigskip
\bf{1} Institute for Emerging Infections, Oxford Martin School, University of Oxford , Oxford, United Kingdom
\\
\bf{2} Nuffield Department of Medicine, University of Oxford, Oxford, United Kingdom
\\
\bf{3} Wellcome Trust Centre for Human Genetics, University of Oxford, Oxford, United Kingdom
\\
\bigskip

* E-mail: nicola.demaio@ndm.ox.ac.uk 

\end{flushleft}

\section*{Abstract}

Exploiting pathogen genomes to reconstruct transmission represents a powerful tool in the fight against infectious disease.
However, their interpretation rests on a number of simplifying assumptions that regularly ignore important complexities of real data, in particular within-host evolution and non-sampled patients.

Here we propose a new approach to transmission inference called SCOTTI (Structured COalescent Transmission Tree Inference). 
This method is based on a statistical framework that models each host as a distinct population, and transmissions between hosts as migration events.
Our computationally efficient implementation of this model enables the inference of host-to-host transmission while accommodating within-host evolution and non-sampled hosts. 
SCOTTI is distributed as an open source package for the phylogenetic software BEAST2.

We show that SCOTTI can generally infer transmission events even in the presence of considerable within-host variation, can account for the uncertainty associated with the possible presence of non-sampled hosts, and can efficiently use data from multiple samples of the same host, although there is some reduction in accuracy when samples are collected very close to the infection time.

We illustrate the features of our approach by investigating transmission from genetic and epidemiological data in a Foot and Mouth Disease Virus (FMDV) veterinary outbreak in England and a \emph{Klebsiella pneumoniae} outbreak in a Nepali neonatal unit. Transmission histories inferred with SCOTTI will be important in devising effective measures to prevent and halt transmission.

\section*{Author Summary}

We present a new tool, SCOTTI, to efficiently reconstruct transmission events within outbreaks.
Our approach combines genetic information from infection samples with epidemiological information of patient exposure to infection.
While epidemiological information has been traditionally used to understand who infected whom in an outbreak, detailed genetic information is increasingly becoming available with the steady progress of sequencing technologies.
However, many complications, if unaccounted for, can affect the accuracy with which the transmission history is reconstructed.
SCOTTI efficiently accounts for several complications, in particular within-patient genetic variation of the infectious organism, and non-sampled patients (such as asymptomatic patients).
Thanks to these features, SCOTTI provides accurate reconstructions of transmission in complex scenarios, which will be important in finding and limiting the sources and routes of transmission, preventing the spread of infectious disease.


\section*{Introduction}

Understanding the dynamics of transmission is fundamental for devising effective policies and practical measures that limit the spread of infectious diseases.
In recent years, the introduction of affordable whole genome sequencing has provided unprecedented detail on the relatedness of pathogen samples~\cite{DIDEETAL12b, WILS12, KOSEETAL12, LEDIEP13}. As a result, inferring transmission between hosts with accuracy is becoming more and more feasible.
However, this requires robust, and computationally efficient methods to infer past transmission events using genetic information.
Many complications, such as within-host pathogen genetic variation and non-sampled hosts, obscure the relationship between pathogen phylogenies and the history of transmission events, affecting the accuracy of such methods.
Here, we present a new approach, SCOTTI, that accounts for these complexities in a computationally feasible manner.  

\begin{figure}
\caption{{\bf Examples of transmission complexities.}
Reconstruction of transmission can be hindered by several complexities causing disagreement between the actual transmission history and the phylogeny of the sampled pathogen.
Here we show four examples of these complexities: \textbf{A}) Within-host evolution (the incomplete lineage sorting problem), \textbf{B}) Incomplete transmission bottlenecks (or large transmission inocula), \textbf{C}) Non-sampled hosts (such as unknown or asymptomatic hosts), \textbf{D}) Multiple infections of the same host (or mixed infections).
Different hosts (named H1, H2, and H3) are represented as black rectangles, and the rectangle with a dashed border represents a non-sampled host (a host for which no pathogen sample has been collected and sequenced, and for which there is no exposure time information). 
The top and bottom edge of each rectangle indicate the introduction and removal times, that is, the beginning and the end of the time interval within which a host is either infective or can be infected (e.g., arrival and departure time from the contaminated ward).
Red dots represent pathogen sequence samples (respectively S1, S2, and S3), and red lines are lineages of the pathogen phylogeny. 
Blue tubes represent transmission/bottleneck events, where the contained lineages are transferred between hosts.
Below each ``nested" tree plot (representing phylogeny and transmission tree simultaneously, see Figure \ref{exampleTrees} in \nameref{S1_Text}), the corresponding transmission history is represented with black ``beanbags", and, in red, the phylogenetic tree of the sequences.}
\hspace{-0.cm}\includegraphics[width=0.9\textwidth]{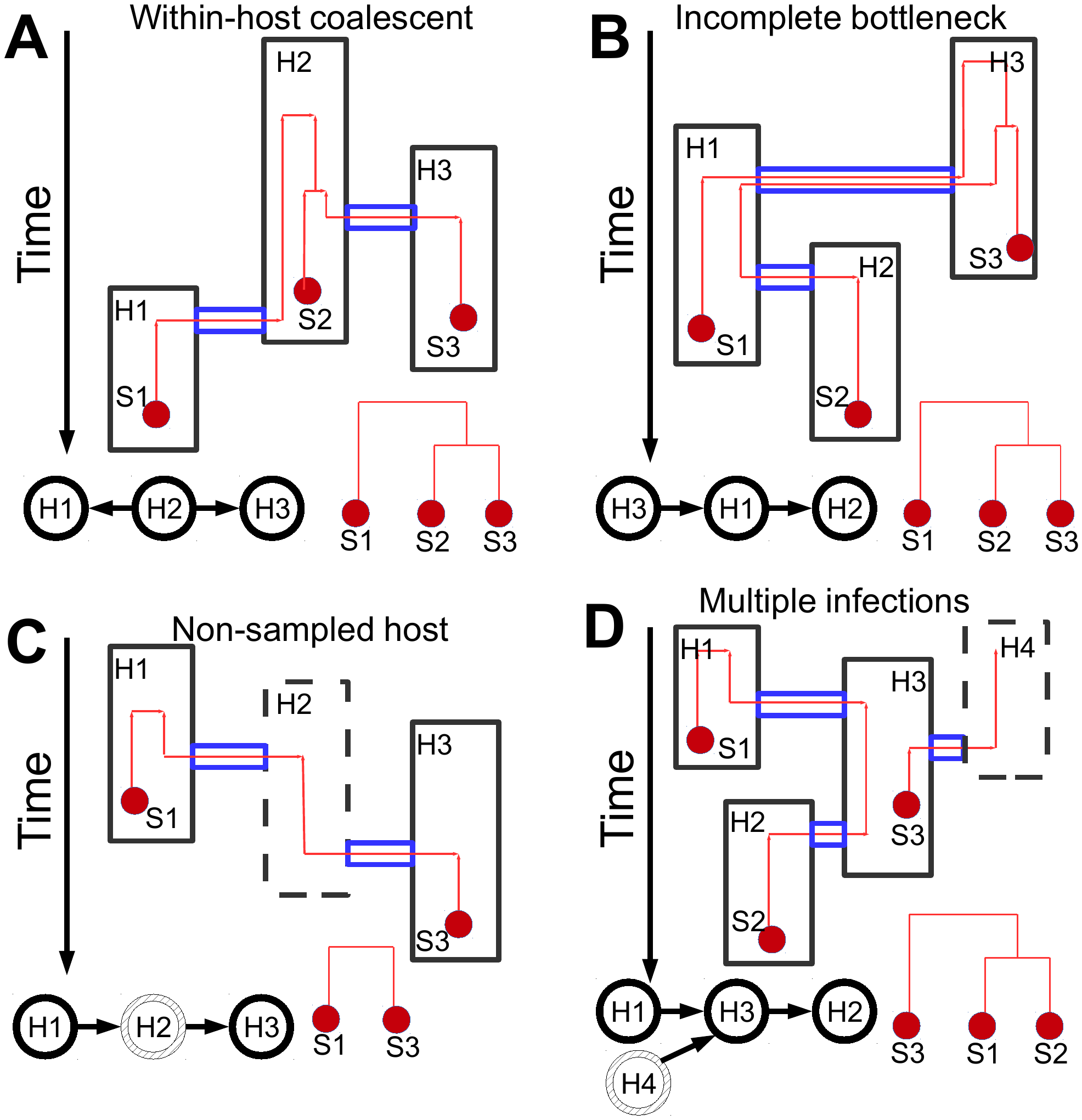}
\label{examples}
\end{figure}

A number of approaches have been developed that reconstruct transmission from genetic data.
One method, based on pathogen genetic data, rules out direct transmission if isolates from different hosts are separated by a number of substitutions above 
a fixed threshold~\cite{EYREETAL13b, WALKETAL13, WALKETAL14}. 
This approach cannot generally distinguish direct transmission from transmission through one or more intermediate hosts, or infer its direction.
Alternatively, the phylogenetic tree of the pathogen samples is often used as a proxy for the transmission history~\cite{LEITETAL96, HARRETAL10}.
While phylogenetic signal can be very informative of transmission, it can also be misleading~\cite{PYBURAMB09, WORBETAL14}.
The main cause of this problem is within-host variation that can generate discrepancies between the phylogenetic and epidemiological relatedness of hosts, and can bias estimates of infection times~\cite{ROMEETAL14}.
One problem arising from within-host diversity is that the pathogen isolates transmitted by a host are not necessarily genetically identical to those sampled from the same host.
This phenomenon can be mathematically modelled using population genetics approaches such as the coalescent~\cite{KING82}, to describe within-host evolution (Figure~\ref{examples}A).
Other factors that can cause disagreement between phylogeny and transmission history are: (i) Incomplete transmission bottlenecks, where some of the within-host genetic variation is transmitted from donor to recipient through a non-negligibly small inoculum; this means lineages from the same host may not have shared a common ancestor since long before the time of infection of the host (Figure~\ref{examples}B). (ii) Non-sampled hosts, such that a sampled patient is not necessarily linked by direct transmission to its most closely related sampled patient, but can have a non-sampled intermediate (Figure~\ref{examples}C)~\cite{VOLZFROS13}. (iii) Multiply infected hosts, that can cause patients to be erroneously excluded from some transmission chains, in particular if multiple samples from the same patient are not collected (Figure~\ref{examples}D).

Several methods emerged in recent years explicitly modelling both the transmission process and genetic evolution to perform inference of the history of transmission events~\cite{COTTETAL08, ALDRETAL11, JOMBETAL11, LIEBETAL11, MOREETAL12, YPMAETAL12,YPMAETAL13, VOLZFROS13, WORBETAL14,  JOMBETAL14, MOLLETAL14,DIDEETAL14, HALLRAMB15, ROMEETAL16}.
These methods generally make use of epidemiological dating information (such as the date of sampling, the interval of exposure of a host to an outbreak, or the likely duration of infectiousness), but they usually ignore within-host variation and other causes of phylogenetic discordance with transmission history~\cite{COTTETAL08, ALDRETAL11, JOMBETAL11, LIEBETAL11, MOREETAL12, YPMAETAL12, VOLZFROS13, JOMBETAL14, MOLLETAL14}.
The methods of Ypma and colleagues~\cite{YPMAETAL13}, Didelot and colleagues~\cite{DIDEETAL14}, and Hall and colleagues~\cite{HALLRAMB15} account for within-host diversity, but assume that all hosts in the outbreak have been detected and sequenced, which may be incorrect or uncertain in practical settings.

Here, we propose a new Bayesian approach called SCOTTI (Structured COalescent Transmission Tree Inference) that not only accounts for diversity and evolution within a host, but also for other sources of bias, namely non-sampled hosts and multiple infections of the same host.
This new method builds on our recent progress in efficiently modelling migration between populations using an approximation to the structured coalescent~\cite{DEMAETAL15}.
Formally, we model each host as a separate pathogen population, and we model transmission as migration between hosts.
SCOTTI has a broad range of applicability as it relaxes the typical assumptions that every host is sampled and that there is no within-host variation (see Figure~\ref{Figure1} in \nameref{S1_Text}).
A limitation of our method is that we do not model transmission bottlenecks.
This can be a disadvantage with strong bottlenecks at transmission (due to small inocula), but on the other hand it may be an advantage with large transmission inocula.
SCOTTI is implemented as an open-source package for the Bayesian phylogenetic software BEAST2~\cite{BOUCETAL14}, and as such, it can be freely installed and used.
We compare the performance of SCOTTI and the popular software Outbreaker~\cite{JOMBETAL14} (version 1.1-5) on simulated data and on real datasets of Foot and Mouth Disease Virus (FMDV~\cite{COTTETAL08b}) and \emph{Klebsiella pneumoniae}~\cite{STOEETAL14}.
These applications highlight how the two methods usually provide very different interpretations of outbreak dynamics, with SCOTTI showing typically higher accuracy on simulated data.
By combining epidemiological and genetic information, and by implementing a general and computationally efficient model, SCOTTI can accurately infer transmission in a broad range of settings, providing important information to understand and limit the spread of infectious disease.

\section*{Results}

\subsection*{SCOTTI: A new Approach to Reconstructing Transmission Events}

\begin{figure}
\caption{{\bf Graphical representation of models of transmission and evolution.}
In the present work we consider three different models of pathogen evolution within an outbreak: \textbf{A)} The multispecies coalescent model with transmission bottlenecks, used for simulations, \textbf{B)} The structured coalescent (SCOTTI) model used for inference, \textbf{C)} The Outbreaker model also used for inference.
The pictures highlight some key parameters and features of the models.
Different hosts (H1, H2, H3, and H4) are represented as black rectangles. 
The top and bottom edge of each rectangle are the introduction and removal times of the respective hosts in \textbf{A} and \textbf{B}.
The hosts with a dashed border are non-sampled. 
Red dots represent samples (only one per host allowed by Outbreaker), red vertical lines are lineages of the phylogeny.
Smaller black dots represent coalescent events. 
Red arrows are transmissions/migrations in \textbf{B} and \textbf{C}. 
Blue tubes are transmissions with bottlenecks in \textbf{A}, and transmitted lineages are contained within them.
In \textbf{A}, a transmission bottleneck from host H1 to H2 causes two lineages in H2 to coalesce (find a common ancestor backwards in time) at the same time of transmission.
This does not happen at the transmission from H3 to H4, where the two lineages in H4 do not coalesce (incomplete bottleneck) and are both inherited from H3 to H4 at a single transmission event.}
\center
\vspace{-0.4cm}
\includegraphics[width=0.99\textwidth]{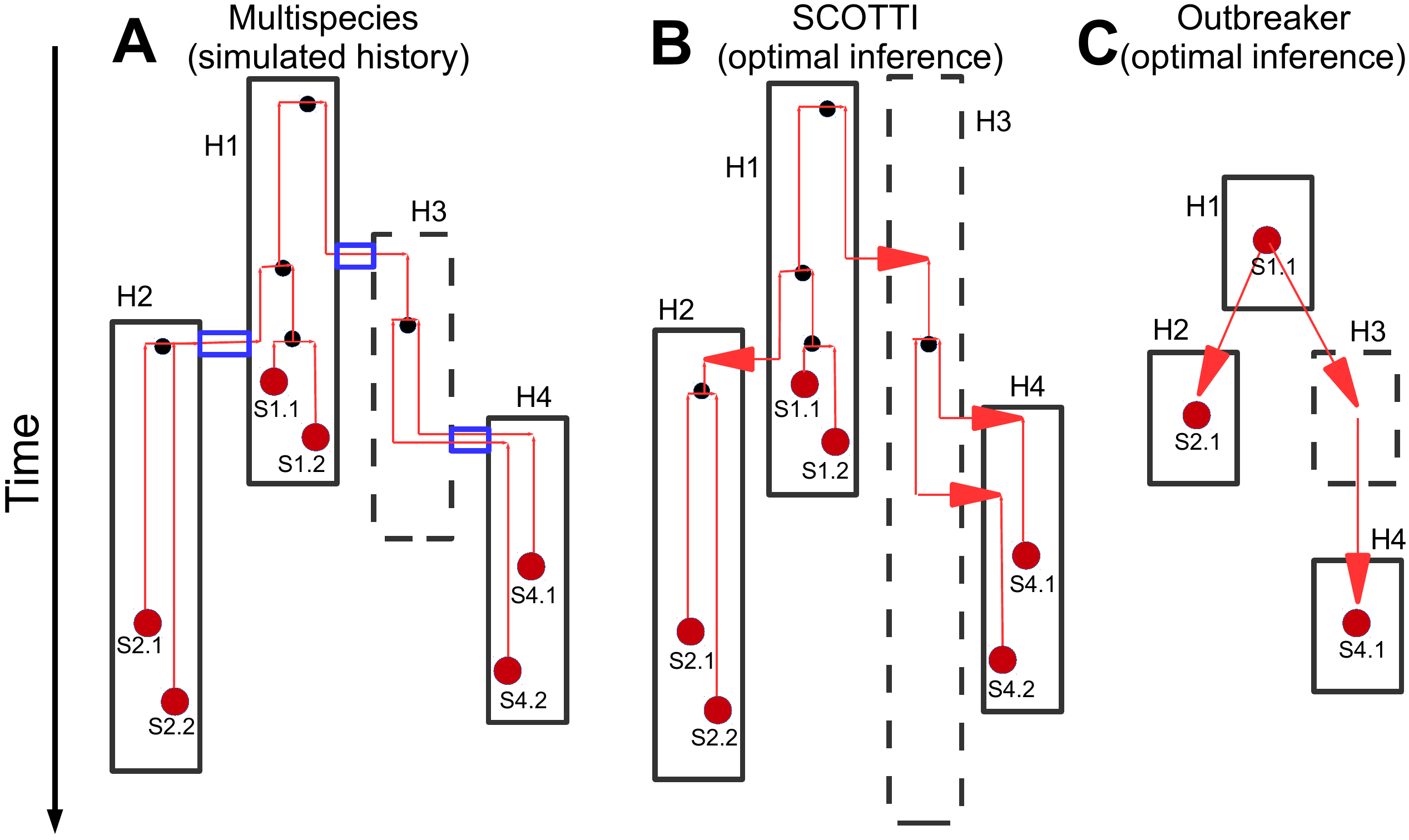}
\label{Figure2}
\end{figure}

Many methods that infer transmission from pathogen genetic data assume that the pathogen population within a host is genetically homogeneous, thereby overlooking within-host variation.
A popular example is Outbreaker~\cite{JOMBETAL14}, where pathogen genetic mutations are assumed to happen during transmission between hosts, and not within hosts (Figure~\ref{Figure2}C).
The simplicity of this model allows estimation of transmission events for pathogens with short and regular incubation and recovery times, and with negligible within-host variation.
However, within-host evolution and overlapping infection intervals are often not negligible for most pathogens, particularly for bacterial infections and chronic viral infections.
If unaccounted for, these complexities can lead to misleading inference concerning transmission events~\cite{WORBETAL14, ROMEETAL14}.

A natural way to account for within-host evolution is via an extension of the multi-species coalescent model~\cite{RANNYANG03} including transmission bottlenecks.
In this model, each host constitutes a separate pathogen population, and with a transmission event some isolates are transmitted to, and colonize, a new host, instantaneously growing to a full and constant-size pathogen population (see e.g.~\cite{DIDEETAL14} and Figure~\ref{Figure2}A).
While this multispecies model is advantageous in several respects, it is usually too computationally demanding to be useful for inference.
So while we base SCOTTI on a simplified model, we simulate pathogen evolution (to assess and compare methods performance) under the multispecies model. 
In these simulations, we fix the host-to-host transmission history, and simulate evolution of pathogen lineages within hosts.
We use two distinct transmission histories inferred from the literature~\cite{COTTETAL08,LEITETAL96}.
We simulate under a broad range of scenarios: different transmission bottleneck severities (weak vs. strong), one vs. two samples per host, different numbers of non-sampled hosts, different amounts of genetic information, and different times of sampling (early, vs. late, vs. randomly within a host infection).
We give further details on the simulation scenarios in the Materials and Methods.

While we simulate outbreak data under the multispecies model, to infer transmission we propose a model based on the structured coalescent, SCOTTI.
In the structured coalescent multiple distinct populations are present at the same time, lineages in the same population can coalesce (find a common ancestor), and lineages can migrate between populations at certain rates.
In SCOTTI, each host represents a distinct pathogen population, and migration of a lineage represents a transmission event (Figure~\ref{Figure2}B).
Lineages are only allowed to evolve within, and migrate to, hosts that are exposed at a given time, and exposure times are informed by epidemiological data.
Under this model, we perform estimation using a new implementation of BASTA (BAyesian STructured coalescent Approximation), an efficient approximation to the structured coalescent~\cite{DEMAETAL15}, adapted to this epidemiological setting.
The use of the approximations in BASTA substantially reduces computational demand, in particular when many populations are considered.
More details on SCOTTI are provided in the Materials and Methods.

\subsection*{Accuracy of Inference on Simulated Data}

\begin{figure}
\caption{{\bf Accuracy of SCOTTI vs. Outbreaker in the base simulation scenario.}
In our base simulation setting, SCOTTI has higher accuracy than Outbreaker, in particular when provided multiple samples per host.
The coloured ``Maypole" tree (see Figure \ref{exampleTrees} in \nameref{S1_Text}) represents the first transmission history used for simulations, with one colour associated to each host, internal nodes corresponding to infection events and times, and tips representing infection clearance times.
The pie charts refer to the accuracy of transmission estimation in the base scenario with strong bottleneck.
The coloured slice in each pie chart is the proportion of replicates (out of a total of 100) for which the correct origin of transmission has been correctly inferred.
Pie charts are plotted below the branch corresponding to the transmission they refer to, while the pie charts for the index host K are plotted next to the root.}
\hspace{-0.0cm}\includegraphics[width=0.98\textwidth]{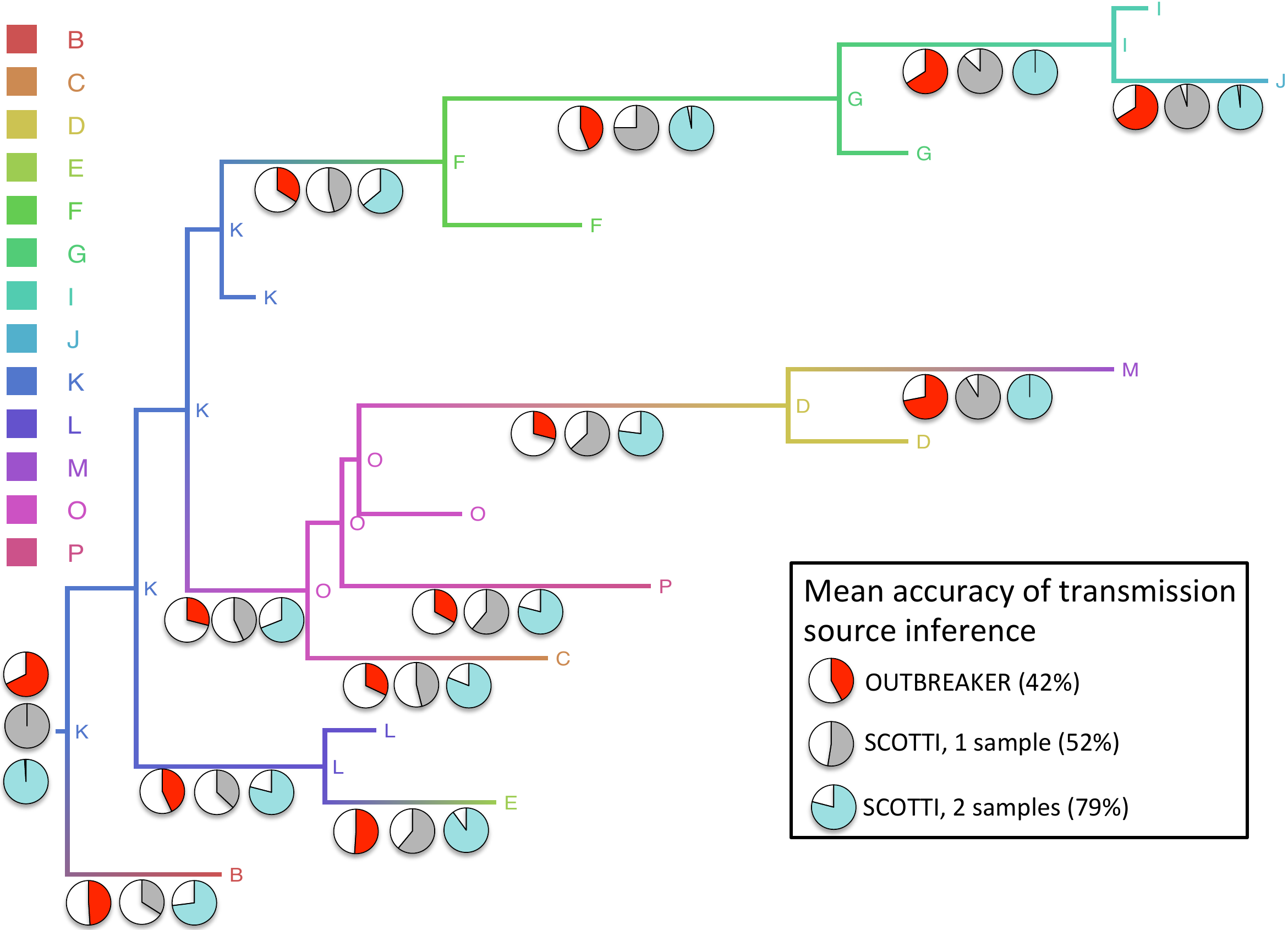}
\label{Base_simple_new}
\end{figure}

\begin{figure}
\caption{{\bf Summary of transmission inference accuracy.}
SCOTTI shows higher accuracy than Outbreaker in all scenarios except with early sampling, while Outbreaker credible sets are poorly calibrated.
Pathogen sequence evolution was simulated under transmission history 1, used in \textbf{A} and \textbf{C}, and transmission history 2, used in \textbf{B} and \textbf{D}. 
 In \textbf{A} and \textbf{B} bars represent proportions, expressed as percentages, of correct inferences of transmission origin (i.e. donor host) over 100 replicates and all transmission events for each method (differentiated by colour as in legend).
On the X axis are different simulation scenarios.
In \textbf{C} and \textbf{D} bars represent average posterior supports, again expressed as percentages, for the correct sources over all patients and replicates.
In \textbf{E} and \textbf{F} bars represent proportions (expressed as percentages) of $95\%$ posterior credible sets that contain the simulated (true) origin. 
The $95\%$ posterior credible set for a host is the minimum set of origins with cumulative probability $\geq 95\%$, and such that all origins in the set have higher posterior probability than all origins outside of it.}
\hspace{-2.0cm}\includegraphics[width=1.19\textwidth]{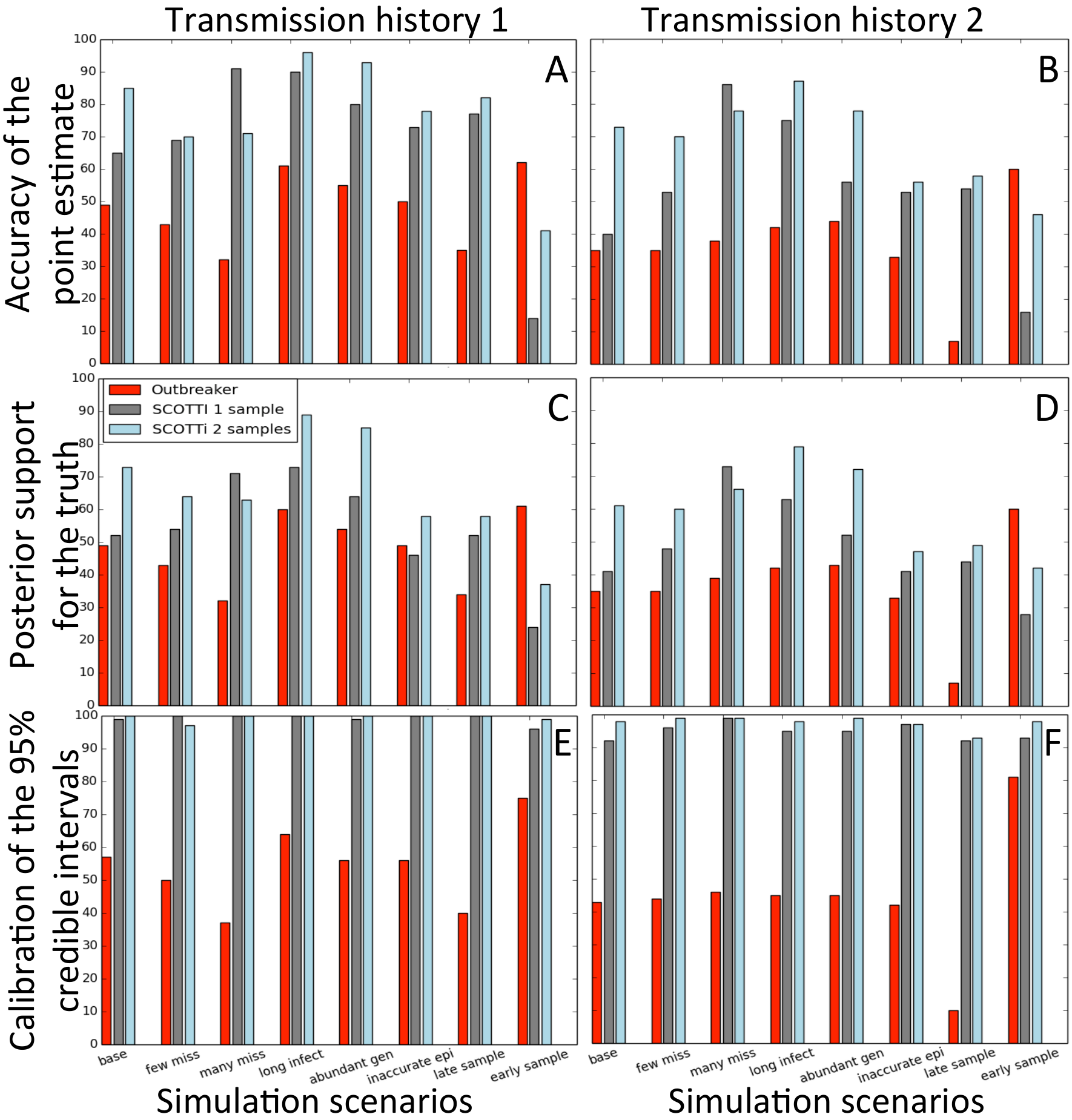}
\label{Summary}
\end{figure}

To test the accuracy of our new method SCOTTI in inferring the origin of transmission, and to compare it to the accuracy of the software Outbreaker, we simulated pathogen evolution within two distinct, fixed transmission histories.
We measure the accuracy of the method as the frequency of correct point estimates of transmission source. As the point estimate we take the transmission donor with the highest posterior probability.
In the base simulation scenario (random sampling times, low genetic variation, every host sampled) SCOTTI performs well in the first transmission history ($65\%$ and $85\%$ mean accuracy with respectively one or two samples per host), and less well on the second transmission history ($40\%$ and $73\%$ accuracy, see Figure~\ref{Base_simple_new} and Figure~\ref{Base} in \nameref{S1_Text}).
In fact, while in the first transmission history transmission events are quite homogeneously distributed through time, in the second history they happen very early, with samples collected much later.
Bottleneck size seems to have a limited effect on inference, although we did not simulate extremely weak bottleneck.
Looking at all other simulation scenarios, we observe that the accuracy of SCOTTI remains consistently high, with the noticeable exception of the case in which sampling occurs very early in infection (Figure~\ref{Summary}).
One likely reason for this is that SCOTTI does not model transmission bottlenecks.
With early sampling, coalescent events are likely to happen in the limited time interval between infection (corresponding to the transmission bottlenecks) and sampling.
Because it is not constrained, SCOTTI often places such coalescent events in the wrong host.

Overall, SCOTTI shows higher accuracy than Outbreaker across scenarios (Figure~\ref{Summary}).
Generally Outbreaker has poor accuracy in estimating the source of transmission for most links in our base scenario, with an average of $49\%$ in transmission history 1 and $35\%$ in transmission history 2 (Figure~\ref{Base_simple_new} and Figure~\ref{Base} in \nameref{S1_Text}).
Its limited performance can be largely explained by the fact that Outbreaker does not account for within-host variation.
This showcases the utility of an approach of broader applicability such as SCOTTI.
SCOTTI generally outperforms Outbreaker even with a single sample per host, and as more samples are included, it achieves greater accuracy ($85\%$ and $73\%$ mean accuracy, vs. $49\%$ and $35\%$ of Outbreaker).
The only instance of inaccuracy of SCOTTI is the transmission from host P5 to host P6 (Figure~\ref{Base} in \nameref{S1_Text}), probably due to the limited phylogenetic evidence supporting it. 
Outbreaker, on the other hand, shows acceptable accuracy when the order of transmission is largely reflected by the order of sampling, and low accuracy otherwise, as for example for the transmissions from host P1 to host P8 and from host P1 to host P5 (Figure~\ref{Base} in \nameref{S1_Text}).

Another difference between the two methods is that Outbreaker tends to infer a posterior distribution supporting a narrower range of origins. This, paired with its limited accuracy, leads the method to exclude a true origin from $95\%$ credible sets in about half of the simulations.
SCOTTI is instead much better calibrated, with $95\%$ credible sets containing the true origin between $90\%$ and $100\%$ of the time (Figure~\ref{Summary} E and F).
Also, most of the inaccurate inference of SCOTTI derives from assigning the source of transmission to non-sampled hosts, i.e., SCOTTI is tentative in naming sampled hosts as transmission donors.
While it is possible to inform SCOTTI of the absence of non-sampled hosts by specifying a strong prior on the corresponding parameter, we prefer not to do so, since in general this information is not available for real outbreaks.
In Outbreaker, inference errors often involve misattributing the source of infection source to one of the sampled hosts (Figure~\ref{SummaryError} in \nameref{S1_Text}).
This considerably affects estimation when genetic and epidemiological data from some hosts is withheld (one host in the ``few missing" scenario and three hosts in the ``many missing" scenario).
In these settings the accuracy of SCOTTI often increases (as non-sampled hosts are correctly attributed to be the source), while it decreases for Oubreaker (since infection source is wrongly attributed to sampled hosts, see Figure~\ref{Missing} in \nameref{S1_Text}).

In most of our scenarios the amount of genetic information available to distinguish different transmission histories is rather limited, with 2-3 SNPs per sampled host on average.
From such data a single phylogenetic tree relating the sequenced samples cannot be inferred unambiguously.
When we increase phylogenetic signal, either by simulating longer infection times (``long infection" scenario) or with longer genetic sequences (``abundant genetic"), the accuracy of the methods substantially increases (Figure~\ref{MoreData} in \nameref{S1_Text}).

SCOTTI and Outbreaker require distinct formats for epidemiological information.
Outbreaker requires as input a probability distribution over the possible durations and intensity of infectivity, and sampling times.
In contrast, SCOTTI requires the user to specify an exposure interval for each host.
In simulations where the exposure intervals provided for each host were doubled in length compared to the true ones (``inaccurate epi" scenario) SCOTTI appeared relatively robust (Figure~\ref{Summary} and Figure~\ref{Inaccurate} in \nameref{S1_Text}).

Furthermore, we investigated the effect of sampling times on the two methods.
Outbreaker has higher accuracy when sampling times are close to the start of infection (``early sampling" scenario). Indeed this is the one setting in which Outbreaker outperforms SCOTTI in inference accuracy (Figure~\ref{Summary}).
In this case SCOTTI is overly tentative in identifying sampled hosts as the source of transmission, and is overly conservative in its quantification of uncertainty (as in the second transmission tree with two samples per host, Figure~\ref{SummaryError} in \nameref{S1_Text}). A contributory factor to this behaviour is that SCOTTI does not model transmission bottlenecks, and so a sampled lineage is not readily inferred to have coalesced (to have found a common ancestor with another lineage) in the short time between sampling and infection, and therefore infers too great a contribution of non-sampled hosts. In situations where a single host transmits multiple times in close succession (as patient P1 in the second transmission history), lineages from its recipients will tend to coalesce within P1 in random order before coalescing with the samples from P1. This means that phylogenetic trees are relatively uninformative of the transmission history, and this inflates uncertainty in the identification of the transmission source.  
In contrast, Outbreaker is less accurate when sampling times are close to clearance times (``late sampling" scenario), see Figure~\ref{Sampling} in \nameref{S1_Text}.

\subsection*{Analysis of FMDV and \emph{Klebsiella pneumoniae} outbreaks}

To investigate the impact of our method on the study of real outbreaks, we examined the transmissions inferred by SCOTTI and Outbreaker in two real outbreaks of FMDV in 2007~\cite{COTTETAL08b} and \emph{K. pneumoniae} in 2011-2012~\cite{STOEETAL14}.

\begin{figure}
\caption{{\bf Reconstruction of transmission events in a FMDV outbreak.}
Outbreaker (\textbf{A}) and SCOTTI (\textbf{B}) provide different interpretations of the 2007 South of England FMDV outbreak.
\textbf{A)} ``Beanbag" tree (see Figure \ref{exampleTrees} in \nameref{S1_Text}) of Transmission events inferred with Outbreaker. The two numbers on each transmission arrow represent respectively the number of nucleotide substitutions separating two hosts, and the inferred posterior support of the event (in this case always $1$,  meaning $100\%$ support). All transmissions are inferred to be direct with more than $95\%$ posterior probability.
\textbf{B)} ``Beanbag" tree of transmission events inferred with SCOTTI. Numbers within host circles represent the posterior probabilities of the corresponding host being the index host (the root) of the considered outbreak. Numbers on arrows represent the inferred posterior probabilities of the corresponding direct transmission events. Colour intensity is proportional to posterior probability.}
\hspace{-6.1cm}\includegraphics[width=1.49\textwidth]{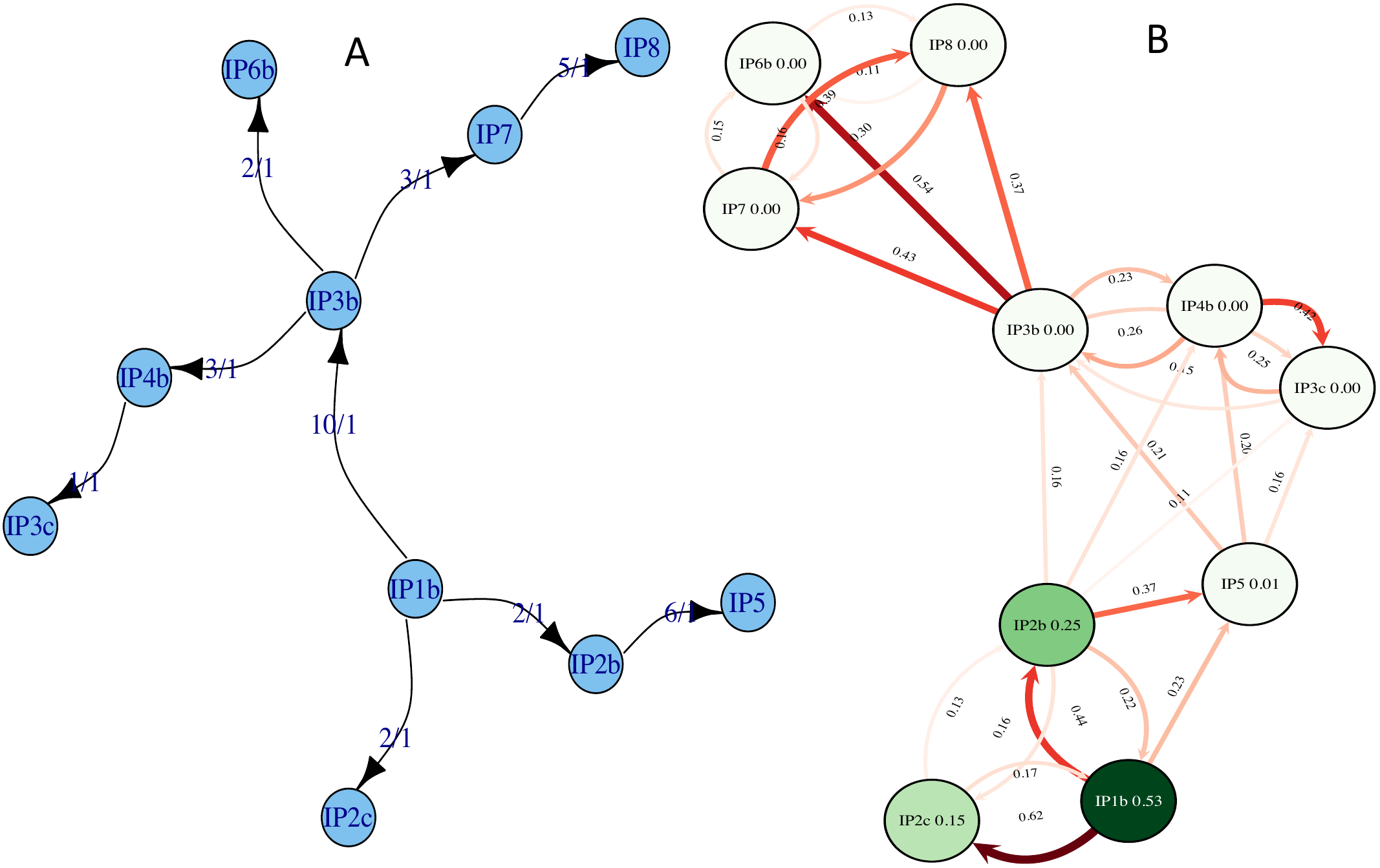}
\label{realFMDV}
\end{figure}

FMDV infects cloven-hoofed animals, and is an economically devastating disease for the farming sector. The 2007 FMDV outbreak occurred in the South England as two distinct transmission clusters, one in August and one in September (Figure \ref{FMDV_Epi} in \nameref{S1_Text}), at an estimated cost to the economy of more than 100 million pounds ~\cite{COTTETAL08b}. Among the questions facing investigators were the source of the outbreak, and the connection between the August and September clusters. Following previous investigations of this outbreak, we studied transmission at the farm-to-farm level, rather than at the scale of individual animals, by taking farms or other geographically delimited premises as the unit of transmission.
Based on previously published whole genome FMDV data ~\cite{COTTETAL08b}, Outbreaker estimated transmission events with high certainty ($100\%$ posterior probability, Figure \ref{realFMDV}A). Yet, some of these inferred events are inconsistent with exposure data and with the transmission events inferred in~\cite{COTTETAL08b} using genetic and epidemiological data (Figure \ref{FMDV_Epi} in \nameref{S1_Text}).
For example, Outbreaker infers $IP1b\rightarrow IP3b$ and $IP3b\rightarrow IP4b$ instead of $IP4b\rightarrow IP3b$ and $IP5\rightarrow IP4b$~\cite{COTTETAL08b}.
This is in part because Outbreaker does not make use of host-specific exposure data, and that its model of genetic evolution does not account for the fact that a host sampling time can be distant from its infection time, as is probably the case for host IP5 here, which was possibly subject to infection for a longer time than other hosts. 
SCOTTI instead considers a much broader range of possible transmission events (Figure~\ref{realFMDV}B). 
The transmission origins with highest posterior probability inferred by SCOTTI correspond to those inferred in~\cite{COTTETAL08b}, consistent with the fact that both methods use exposure data.
Another reason to believe that SCOTTI is more reliable in this case is the sampling scheme, which is very close to our simulated ``late sampling'' scenario, where we observed SCOTTI to be more accurate (Figure~\ref{Summary}).
As shown in simulations, much of the uncertainty in SCOTTI is attributed to the possible presence of non-sampled hosts, which could be reduced by modifying the prior on the number of non-sampled hosts.
Regarding the connection between the August and the September clusters, Cottam and colleagues identified IP5 as a possible link in the transmission chain, but did not exclude the possibility of alternative, transient, unobserved infections between the two clusters.
SCOTTI adds weight to the notion of unobserved and non-sampled intermediate infections; the posterior probability of non-sampled intermediates was $\approx 66\%$, suggesting that it is likely that some important transmission links might not have been sampled.

\begin{figure}
\caption{{\bf  Reconstruction of Transmission events in a \emph{K. pneumoniae} outbreak.}
Outbreaker (\textbf{A}) and SCOTTI (\textbf{B} and \textbf{C}) provide different interpretations of the \emph{K. pneumoniae} outbreak.
\textbf{A)} ``Beanbag" tree of transmission events inferred with Outbreaker. Each circle represents a host, with ``PMK" removed from their name. The number on transmission arrows represents the inferred posterior probability of the event. All arrows represent direct transmissions (without intermediate non-sampled hosts, with more than $85\%$ support) except the one from PMK9 to PMK10 which is inferred to be through at least one intermediate host.
\textbf{B)} ``Beanbag" tree of transmission events inferred with SCOTTI. Numbers on arrows represent the inferred posterior probabilities of the corresponding direct transmission events. Colour intensity is proportional to posterior support.
\textbf{C)} ``Maypole" maximum clade credibility tree (see Figure \ref{exampleTrees} in \nameref{S1_Text}) inferred with SCOTTI, annotated and coloured with the highest posterior probability hosts for internal nodes. ``NS" represents all non-sampled hosts. Branch width indicates the posterior  probability of the inferred host at the node at the right end of the considered branch. Branches are annotated with $95\%$ posterior intervals of the number of transmissions. For non-annotated branches, the interval is $[0,1]$.}
\hspace{-6.8cm}\includegraphics[width=1.59\textwidth]{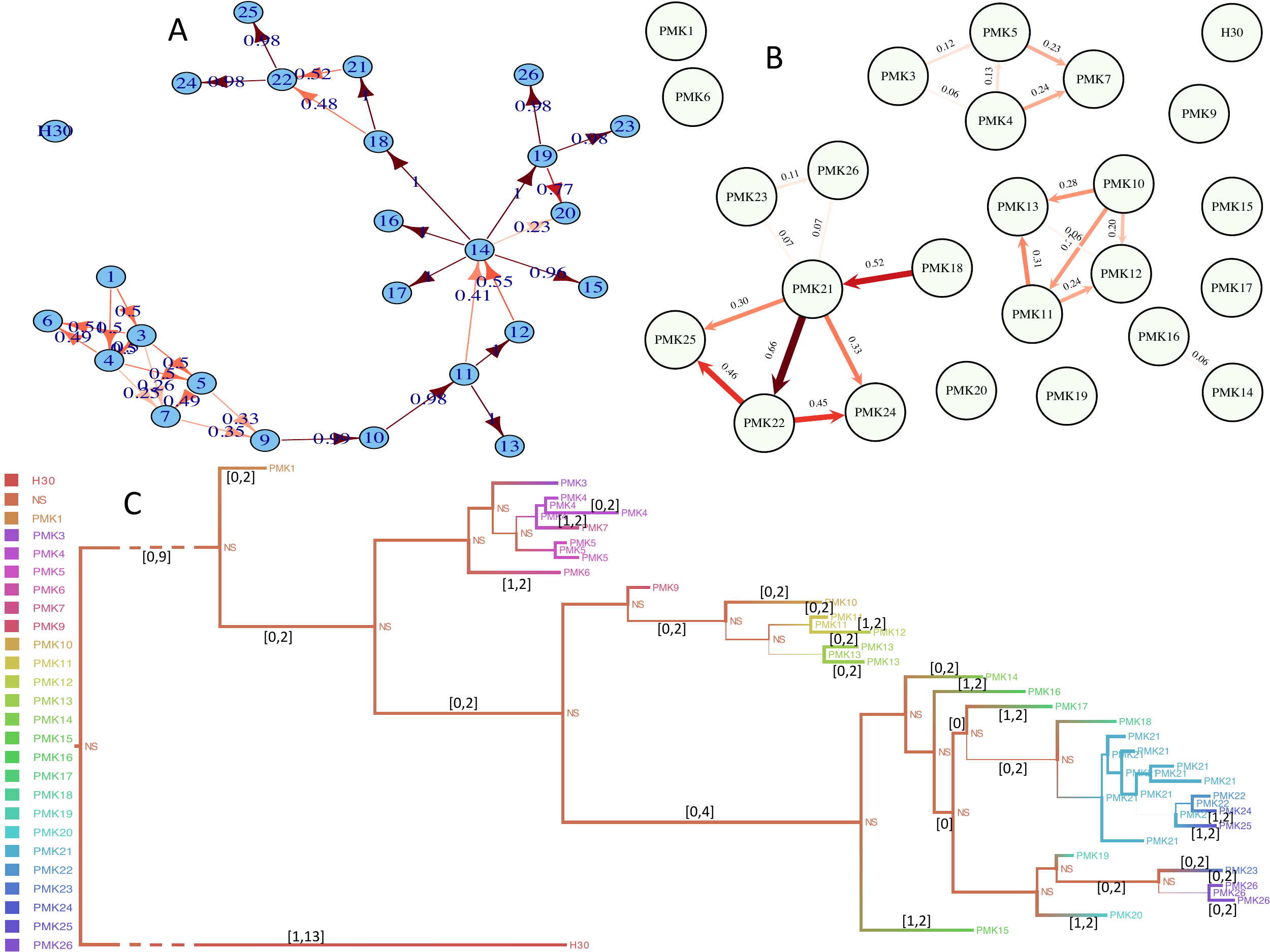}
\label{realKPneu}
\end{figure}

Secondly, we investigated an antimicrobial resistant \emph{K. pneumoniae} outbreak in a Nepali neonatal intensive care unit between August 2011 and June 2012. \emph{K. pneumoniae} antimicrobial resistant strains are a major health concern particularly for neonatal clinical care. In this outbreak, there were 16 neonate deaths out of 25 infections, representing a very high case fatality rate (64$\%$) ~\cite{STOEETAL14}. Of major importance to the outbreak investigation was the role of transmission within the unit, vs recurrent introduction from outside.
In this outbreak, Outbreaker inferred just one transmission event to be indirect (through a non-sampled host, from host PMK9 to PMK10, with probability $\approx 91\%$).
All other transmission events were inferred to be between sampled hosts with posterior probability above $85\%$ (although this probability was often distributed over multiple possible sampled sources). 
Two patients were inferred to represent novel introductions (index cases): PMK1 and H30.
While most infections were attributed to a single patient with greater than $99\%$ probability, the source of infection of many patients was considerably uncertain (PMK3-9, PMK14, PMK20 and PMK22, see Figure~\ref{realKPneu}A).
SCOTTI, in complete contrast, inferred a non-sampled source as the most likely for the majority of sampled patients (Figure~\ref{realKPneu}B).
Direct transmission between sampled hosts was only inferred with high confidence for a small number of pairs, the most likely being $PMK18\rightarrow PMK21$, $PMK22\rightarrow PMK24$, $PMK22\rightarrow PMK25$.
Overall, SCOTTI inferred sampled patients to constitute a small portion of the total outbreak, forming separated (Figure~\ref{realKPneu}B), and yet related (Figure~\ref{realKPneu}C), sub-outbreaks (respectively PMK3-7, PMK9-13, PMK14-26, and the two relatively isolated cases PMK1 and H30) within a larger outbreak.
This is consistent with, and informed by, the presence of four time intervals not covered by any host exposure, requiring the presence of non-sampled infected intermediate hosts, recurrent introductions, or environmental contamination~\cite{STOEETAL14} (Figure \ref{KlebEpi} in \nameref{S1_Text}).
SCOTTI also inferred most of the common ancestors of the sampled patients to be non-sampled (Figure~\ref{realKPneu}C).
This conclusion cannot be reached by Outbreaker, which assumes that the most recent common ancestor of two sampled cases (within outbreaks with a single index case) is also sampled.

\section*{Discussion}

Methods to infer transmission events within outbreaks are essential to determine the causes and patterns of transmission, and therefore to inform policies preventing and limiting transmission.
Genomic data from pathogen samples give the opportunity to investigate, at an unprecedented level of detail, the relatedness of pathogens from different hosts.
However, common real life complexities such as within-host variation (in particular for bacterial and chronic viral infections~\cite{PYBURAMB09, WORBETAL14,ROMEETAL14}), or hosts that have not been sampled (e.g. unknown or asymptomatic patients) can hinder the reconstruction of transmission events.
Therefore, methods that efficiently infer transmission from genomic data while accounting for within-host variation and non-sampled hosts are essential if we want to determine specific transmission events within outbreaks, or even general patterns of transmission which might inform policies and recommendations for prevention of infection.

Here we have presented SCOTTI, a novel method of host-to-host transmission inference (who infected whom) that is built around a computationally efficient model of pathogen evolution based on the structured coalescent.
By modelling each host as a distinct pathogen population, and transmission as migration of lineages between hosts, we have shown that it is possible to model within-host evolution and estimate transmission events with good accuracy, even in the presence of non-sampled hosts.
We compared the accuracy of SCOTTI with that of the similar software Outbreaker~\cite{JOMBETAL14} in a broad range of simulation scenarios: different epidemics (FMDV or HIV), transmission bottleneck sizes, numbers of samples per host, numbers of non-sampled hosts, genome sizes, SNPs per transmission, and sampling times.
Overall, SCOTTI has better accuracy, in particular when benefitting from multiple samples from each host, which Outbreaker does not allow.
SCOTTI, in fact, explicitly models within-host evolution, and multiple samples from the same host can be particularly informative regarding the within-host coalescent rate, and, in theory, the proportion of mixed infections.
While it is common to sample and sequence only one haplotype from each host, this clearly shows that transmission inference benefits greatly from within-host variation data.
Also, SCOTTI explicitly uses epidemiological data in the form of exposure times of the hosts to inform plausible direct transmission events, or equivalently to rule out impossible direct transmissions due to non-overlapping exposure times.
Inaccurate epidemiological data, and even more so the absence of such data, can consequently lead to a decrease in accuracy.
In our simulations, we have observed that partial inaccuracy in epidemiological data has a modest effect on the accuracy of SCOTTI;
yet, we expect that in the complete absence of exposure time information, differently from Outbreaker, the accuracy of SCOTTI would be more deeply compromised.
Further, SCOTTI explicitly models non-sampled hosts, and this helps inference when the outbreak is only partially sampled, as we show in simulations and in the \emph{K. pneumoniae} outbreak.
We showed in particular that not only do SCOTTI and Outbreaker have different accuracy, but also differ with respect to the inferred contribution of non-sampled hosts.
It is therefore important to carefully select the most appropriate software to investigate transmission, and looking at the degree of concordance of different methods might reveal transmission events of difficult attribution.

Although SCOTTI has broad applicability, it has important limitations to be considered that we will address in future work.
One problem is that SCOTTI ignores transmission bottlenecks, that is, the rapid growth in pathogen population size within a host following transmission.
This might cause biases with strong transmission bottlenecks and early samples, as we showed in the simulation setting involving early sampling times.
On the other hand, due to this feature SCOTTI is likely to suit outbreaks with frequent mixed infections and large transmission inocula (see e.g.~\cite{MURCETAL10}), like many gut and environmental bacteria which have high prevalence and low pathogenicity.
The model could also be extended in the future to allow different types of epidemiological data (such as a prior on the time of each infection, while we now only allow a fixed older bound); or also compartmental epidemiological models~\cite{BRAU08}, and geographic distance or structure of different hosts.

In conclusion, we have presented a new method to reconstruct transmission events, SCOTTI, that addresses the urgent need for software to analyse genomic and epidemiological data while accommodating for incomplete or patchy host sampling, mixed infections, and within-host variation. For these reasons, our method can help to reconstruct transmission histories in a broad range of outbreaks, both bacterial and viral. This information will in turn be essential for devising effective strategies to fight the spread of infectious disease.

\section*{Software Availability}

SCOTTI is distributed as an open source package for the Bayesian phylogenetic software BEAST2. It can be downloaded from https://bitbucket.org/nicofmay/scotti/ or via the BEAUti interface~\cite{DRUMETAL12} of BEAST2.

\section*{Materials and Methods}

\subsection*{Approximate Structured Coalescent Model}

Recently, we proposed a BAyesian STructured coalescent Approximation (BASTA) that uses the structured coalescent framework (also known as the coalescent with migration) to infer migration rates and events between populations~\cite{DEMAETAL15}.
BASTA requires substantially less computational time than the exact structured coalescent, particularly when more than just a few populations are considered, by using approximations similarly to~\cite{VOLZ12,RASMETAL14}.
Here, we use the modelling approximations of BASTA in an epidemiological setting, where we model each host as a distinct pathogen population and transmissions as migration events (see Figure~\ref{Figure2}B).
A list of the symbols used hereafter is given in Table~\ref{Symbols} in \nameref{S1_Text}.

To allow the inclusion of epidemiological data, each population (host) $d\in D$ is associated with an exposure interval limited by an introduction time $d_i \in ( -\infty , +\infty ]$ and a removal time $d_r \in [ -\infty , +\infty )$, with $d_r < d_i$ (we consider time backward as typical in coalescent theory).
The interval $[d_r, d_i ]$ represents the exposure interval for population $d$, outside of which $d$ cannot host any pathogen lineage. 
$d_i$ and $d_r$ represent respectively the times at which first it was possible for the host to have been infected, and last to have been infectious. 
For example, in a nosocomial outbreak, $d_i$ and $d_r$ would represent respectively the time of arrival and departure of host $d$ into and from the infected hospital or ward. 
We assume that $d_i$ and $d_r$ are provided by the user and are therefore hereby treated as auxiliary data (we do not model host exposure, and exposure times are always conditioned on). 
In the worst case scenario where no information on host $d$ exposure is provided, it is assumed that $d$ is exposed for the whole outbreak ($d_i=+\infty$ and $d_r=-\infty$).
We will denote as $\bm{E}$ the collection of exposure times.
The number of populations $n_D$ is not fixed, but is estimated within a range specified by the user.
In the remainder of this work, we will assume that non-sampled demes have unlimited exposure times, but we also provide the option in SCOTTI of specifying regularly distributed introduction and removal times.
Here, $n_D$ does not necessarily correspond to the number of non-sampled intermediate hosts in the outbreak, as each host can be infected multiple times, so a non-sampled host due to its infinite exposure time can model more than one non-sampled intermediate host.
An additional important difference to~\cite{DEMAETAL15} is that we assume that the migration (or infection) rate $m$ is the same between each pair of hosts for the time that they are both exposed.
Also, all demes are assumed to have the same effective population size $N_e$.
This means that we assume that transmission is \emph{a priori} equally likely between any pair of exposed hosts, and that all hosts have equal, and constant, within-host pathogen evolution dynamics.
These assumptions of equal population sizes and migration rates simplify the model and distinguish it from classical structured coalescent methods.
In fact, rather than focusing on estimating differences in migration rates and population sizes as in typical structured coalescent methods, we focus on the inference of migration (that is, transmission) events.
Yet, these assumptions could be relaxed for example to account for geographically structured hosts or for known contact network.
For the time that a host $d$ is not exposed, migration rate into or out of $d$ is 0.

We assume that a set of samples $I$ is provided, where each sample $i\in I$ comes with an aligned sequence $s_i \in S$, a sampling date $t_i \in t_I$, and a sampled host $l_i \in L$.
We allow any number of samples from any host, including none (for non-sampled hosts).
In this study, we assume that the molecular evolution process follows a time-homogeneous and site-homogeneous HKY model~\cite{HASEETAL85}, with parameters \bm{$\mu$}. 
However, it could be as general as allowed by BEAST2.
We denote $T$ the bifurcating tree that elucidates the phylogenetic relationships of the samples, and $M$ the migration history of all lineages (the collection of all migration/transmission events).

To infer the transmission history in a Bayesian statistical framework, we aim to approximate the following joint posterior distribution:
\begin{multline}
P(T,M,n_D,\bm{\mu},m,N_e,|S,t_I,L,\bm{E}) \propto \\
\propto P(S|T,t_I,\bm{\mu}) P(T,M|t_I,L,m,N_e,\bm{E},n_D) P(\bm{\mu},m,N_e,n_D).
\label{MTT}
\end{multline}
The first term on the right hand side is the likelihood of the sequences given the genealogy and substitution model.
It assumes that sequences evolves down the tree according to a continuous time Markov chain and it can be generally calculated with Felsenstein's pruning algorithm~\cite{FELS81}. 
The second term is the probability density of the genealogy and migration history. 
The third term represents the joint prior distribution on the parameters of the nucleotide substitution model and the migration model.
Generally, exploring the space of all possible migration histories is computationally demanding even for moderate numbers of populations~\cite{DEMAETAL15}.
For this reason, we integrate over all migration histories 
by approximating the following posterior distribution as in BASTA~\cite{DEMAETAL15}:
\begin{equation}
P(T,n_D,\bm{\mu},m,N_e|S,t_I,L,\bm{E}) \propto P(S|T,t_I,\bm{\mu}) P(T|t_I,L,m,N_e,\bm{E},n_D) P(\bm{\mu},m,N_e,n_D).
\label{BASTA}
\end{equation}

To approximate $P(T|t_I,L,m,N_e,\bm{E},n_D)$ in Eq~\ref{BASTA}, we consider the probability density of each time interval between successive events (coalescence, sampling, population introduction, or population removal events).
The steps below are very similar to those in BASTA, the main differences being that here we assume equal migration rates and population sizes, and we account for population introductions and removals.
Denoting each interval $A_{i}=[\alpha_{i-1} , \alpha_i]$, where $\alpha_{i}$ is the older event time of $A_i$ and $\alpha_{i-1}$ the more recent, the probability density of interval $A_i$ can be written as

\begin{equation}
L_{i} = \exp \left[-\int_{\alpha_{i-1}}^{\alpha_i} \sum_{d \in D} \dfrac{1}{2}\sum_{l \in \Lambda_i} \sum_{l' \in \Lambda_i, l'\neq l} {P(d_l= d, d_{l'} = d | t)} \dfrac{1}{N_e} dt \right] E_i,
\label{trueRate}
\end{equation}
where $\Lambda_i$ is the set of all extant lineages during interval $A_i$, {$d_l$ is the host to which lineage $l$ belongs, and $P(d_l= d, d_{l'} = d | t)$} is the probability that lineages $l$ and $l'$ are in the same host $d$ at time $t$. $E_i$ is the contribution of the particular event:
\begin{equation}
E_i = \left\{  \begin{array}{ccl} \sum_{d \in D} P_{l,\alpha_i,d} P_{l',\alpha_i,d}   \dfrac{1}{N_e} & \mbox{if it is a coalescence between $l$ and $l'$,} \\  1 & \mbox{otherwise.}  \end{array}  \right.
\end{equation}

To approximate $L_{i}$ we substitute {$P(d_l= d, d_{l'} = d | t)$ with $P(d_l= d | t)P(d_{l'} = d | t)$}, which
corresponds to modelling lineages as migrating independently of each other within an interval between events.
This is an approximation in general, but as shown in~\cite{DEMAETAL15} it has limited effect on estimation. 
 As shorthand, we define $\bm{P}_{l,t}$ to be the vector whose $d$th element is $P_{l,t,d}=P(d_l= d | t)$. 
 Next, we split each interval $A_i$ into two sub-intervals of equal length $A_{i1}=[\alpha_{i-1} , (\alpha_i + \alpha_{i-1})/2]$ and $A_{i2}=[(\alpha_i + \alpha_{i-1})/2 , \alpha_i]$, and replace $\bm{P}_{l,t}$ with $\bm{P}_{l,\alpha_{i-1}}$ for all $t$ in $A_{i1}$ and $\bm{P}_{l,\alpha_i}$ for all $t$ in $A_{i2}$. 
 We also call $\tau_i:=\alpha_i-\alpha_{i-1}$. 
 The approximated probability density contributions of $A_{i1}$ and $A_{i2}$ become:

\begin{equation}
\tilde L_{i1} = \exp \left[-\dfrac{\tau_i}{2} \sum_{d \in D} \dfrac{1}{2} \sum_{l \in \Lambda_i} \sum_{l' \in \Lambda_i, l'\neq l} P_{l,\alpha_{i-1},d} P_{l',\alpha_{i-1},d} \dfrac{1}{N_e} \right] 
\label{rate}
\end{equation}

and 

\begin{equation}
\tilde L_{i2} = \exp \left[-\dfrac{\tau_i}{2} \sum_{d \in D} \dfrac{1}{2} \sum_{l \in \Lambda_i} \sum_{l' \in \Lambda_i, l'\neq l} P_{l,\alpha_{i},d} P_{l',\alpha_{i},d} \dfrac{1}{N_e} \right] E^\prime_i.
\label{rate}
\end{equation}

The probability density of the genealogy under the structured coalescent, integrated over migration histories, is finally approximated as

\begin{equation}
P(T|t_I,L,m,N_e,\bm{E},n_D) \approx \prod_{i} \tilde L_{i1}\tilde L_{i2}.
\end{equation}

The probability distribution of lineages among demes is updated iteratively starting from the most recent event toward the past as
\begin{equation}
P_{l,\alpha_i,d}=P_{l,\alpha_{i-1},d}(\dfrac{1}{D_i} + \dfrac{D_i-1}{D_i}e^{-\tau_i m}) + (1-P_{l,\alpha_{i-1},d})(\dfrac{1}{D_i} - \dfrac{1}{D_i}e^{-\tau_i m}) 
\label{Ps}
\end{equation}
for any host $d$ exposed during the considered interval, where $D_i$ is the number of hosts (sampled or non-sampled) exposed during interval $A_i$.
This comes from the assumption that any lineage migrates away from the current host at total rate $m$, and uniformly towards all other extant hosts.
For a lineage $l$ sampled from deme $d$ at time $t$, $\bm{P}_{l,t}$ is a vector whose $d$th element equals one and all other entries equal zero. 
If lineages $l_1$ and $l_2$ coalesce to an ancestral lineage $l$ at time $t$, then 
\begin{equation}
\bm{P}_{l,t} = \dfrac{ \left(  P_{l_1,t,1} P_{l_2,t,1}  , \ldots , P_{l_1,t,n_D} P_{l_2,t,n_D}  \right) }{  \sum_{d=1}^{n_D}  P_{l_1,t,d} P_{l_2,t,d}  }, 
\label{Ps2}
\end{equation}
which is the normalised entrywise product (element by element product) of the distributions of the coalescing lineages.
If instead $\alpha_i=d_i$ is the introduction time for a deme, all remaining lineages in host $d$ are forced to migrate out of host $d$.
First, $\forall l\in \Lambda_i$ we update the probabilities as in Eq.~\ref{Ps}. 
Then, $\forall l\in \Lambda_i$, $P_{l,\alpha_i,d}$ is set to $0$, and its value is distributed uniformly over all other hosts.
If the considered event is a removal of host $d$, Eq.~\ref{Ps} is used again, and $\forall l\in \Lambda_i$ $P_{l,\alpha_i,d}$ is initiated with the value $0$. 

Samples from the posterior distribution in Eq.~\ref{BASTA} are simulated via a Monte Carlo Markov Chain (MCMC).
For each of the MCMC samples we simulate hosts at the internal nodes of $T$, and numbers of transmission events along branches, using the same technique as in~\cite{DEMAETAL15}:
first, starting from the root of $T$ and progressing towards the tips, a host for each internal node is sampled according to its posterior probability (Eq.~\ref{Ps2}) and conditional on the host sampled at its parent node.
Secondly, the numbers of transmission events on each branch are sampled under a Poisson distribution depending on the migration rate $m$ and conditional on if the previously sampled hosts at the extremities of the considered branch are the same or not.

SCOTTI allows a large number of populations to be investigated, as the assumption of uniformity of migration rates and effective population sizes greatly reduces the computational demand and parameter space compared to~\cite{DEMAETAL15}. 
Also, all non-sampled hosts with the same exposure interval are de facto identical, so usually $n_D$ has no effect on the computational demand of SCOTTI.
Example files and data from the analyses described hereby can be found in Supplementary Dataset S1.

\subsection*{Simulations of Pathogen Evolution}

We test the performance in transmission inference of SCOTTI and Outbreaker using a broad range of simulation scenarios.
We simulate within-outbreak pathogen evolution using the transmission events observed in two example real-life outbreaks.
For half of simulations we use a subset of the FMDV transmission history inferred in~\cite{COTTETAL08} (hereby referred to as ``transmission history 1") including 20 UK farms infected during the 2001 outbreak.
For the other half of the simulations we use the HIV transmission history described in~\cite{LEITETAL96} (hereby referred to as ``transmission history 2") of an outbreak occurred between 1981 and 1983, where a male contracted HIV in 1980 and spread it to six females who subsequently infected two male sexual partners and two children.
A description of the transmission history of both scenarios is depicted in Figure~\ref{Base} in \nameref{S1_Text}.

While we simulate the coalescent process randomly, the transmission process is fixed \emph{a priori}, so that always the same transmission events at the same time are considered, and only within-host evolution of lineages is varied in different replicates.
We use a variant of the multi-species coalescent~\cite{RANNYANG03} which is similar to the model used recently in comparable simulations~\cite{DIDEETAL14, ROMEETAL14}.
The model used for simulations is considerably different to both the SCOTTI and Outbreaker models of pathogen evolution (see Figure~\ref{Figure2}).
Hereby each host, during its time of infection, is modelled as a pathogen population with constant effective population size $N_e$, which is the same for all hosts.
Lineages within a host can freely coalesce back in time as in the standard coalescent.

In addition to within-host evolution, we want to simulate a typical transmission: a small proportion of the pathogen population passed on at transmission (due to limited inoculum size), followed by rapid growth in the recipient (see e.g.~\cite{SALAETAL09, BULLETAL11}).
Therefore, we simulate transmission as a backward in time instantaneous bottleneck in the recipient, followed back in time by the merge of the donor and recipient populations into the donor host.
The bottlenecks simulated can have two effect sizes: either equivalent to the drift of $N_e$ generations (a weak bottleneck through which two lineages have a probability of $\approx 63\%$ of coalescing), or $100\, N_e$ generations (a strong bottleneck through which two lineages almost surely coalesce).
For half of the simulation scenarios we use a weak bottleneck, for the other half a strong one.
In the population merger after the bottleneck all lineages remaining in the recipient host are moved to the donor host.
Transmission bottlenecks are neither modelled in SCOTTI, nor in Outbreaker (Figure~\ref{Figure2}). 

Finally, half of the simulations are performed providing one sample per host, the other half providing two samples per host, although Outbreaker is only used with one sample per host as it is the only permitted scenario.
In summary we have $2\times 2\times 2=8$ groups of simulations: 
\begin{itemize} \itemsep0.pt
\item Weak vs strong bottleneck 
\item First  vs second transmission history 
\item One vs two samples per host.
\end{itemize}

For each of the aforementioned eight groups, eight different scenarios (or subgroups) are simulated, for a total of 64 distinct simulation settings.
We define a basic subgroup (called ``base"), and seven variants, in each of which one aspect of the base subgroup is modified.
In ``base", sampling times are picked uniformly at random and independently within host exposure times, the average time of infection is $2\,N_e$ generations, host is sampled, the alignment length is 1500 bp, and  the epidemiological data provided to SCOTTI is accurate (introduction and removal times correspond to infection and recovery time of hosts).
The seven variant settings are:
\begin{itemize} \itemsep0.pt
\item {\bf Long infection} - the intervals of infection are five times longer (on average $10\, N_e$ generations).
\item {\bf Abundant genetic} - the alignment is 10 times longer (15000 bp). 
\item {\bf Early sampling} - samples are collected very early in infection, that is, $5\%$ of the total infection time after infection.
\item {\bf Late sampling} - samples are collected at recovery time for each host.
\item {\bf Few missing} - one host in the outbreak is not sampled (host O for transmission history 1 and host P5 for transmission history 2).
\item {\bf Many missing} - three hosts are not sampled (O, G, and H for transmission history 1, and P1, P5, and P8 for transmission history 2).
\item {\bf Inaccurate epi} - SCOTTI is provided with an exposure interval that is broader than the interval of infection of each host. 
In particular, if $\bar L_i$ is the average length of infection among hosts, then introduction and removal times are respectively $\bar L_i/2$ earlier than infection and $\bar L_i/2$ later than recovery time.
\end{itemize}
For each of the total 64 subgroups, 100 datasets are simulated under an HKY substitution model~\cite{HASEETAL85} with $\kappa=3$, $10^{-3}$ substitution rate per base per $N_e$ generations, and uniform nucleotide frequencies.

For each simulated dataset we infer transmission with Outbreaker under the HKY substitution model and with $10^6$ MCMC iterations. 
Each analysis is initiated with a random starting tree and with uniform prior infection and sampling probabilities over the maximum observed infection time interval.
We run SCOTTI with an HKY substitution model, between 0 and 2 non-sampled hosts, and $10^6$ MCMC iterations.
For both methods we assess the performance by checking how often the correct origin of infection of each sampled host is recovered, and with what posterior probability.
If transmission from host 1 to host 2 is inferred (either by SCOTTI or Outbreaker), then the infection origin of host 2 is inferred to be host 1.
If an indirect transmission to host 2 is inferred, or host 2 is inferred to be an index host or an imported host, then the infection origin of host 2 is inferred to be non-sampled.
Lastly, if SCOTTI infers multiple origins of the same host, then, if more than one origin is a sampled host, we always consider the inference as wrong; otherwise we consider the only sampled origin.
We use two metrics to define the accuracy of an origin inference: (i) the number of replicates in which the simulated origin is the one with the highest posterior probability (ii) the average posterior probability of the simulated origin across replicates.

\subsection*{Foot and Mouth Disease Virus and \emph{K. Pneumoniae} data}

We apply and compare SCOTTI and Outbreaker on two real datasets: one from the 2007 FMDV outbreak in UK~\cite{COTTETAL08b} and one from a \emph{K. pneumoniae} outbreak in a Nepali hospital in 2011-2012~\cite{STOEETAL14}.
In both cases we use alignments, sampling dates and exposure times as provided in the respective studies.
All hosts have a single sample, except IP1b (two samples) for FMDV and PMK4, PMK5, PMK13, PMK26 (two samples) and PMK21 (five samples) for \emph{K. pneumoniae}.
When multiple samples are present for the same host, only the oldest sample is provided to Outbreaker (which only allows one sample per host).
Outbreaker is run for $2\times 10^6$ MCMC iterations on FMDV, and $10^9$ iterations on \emph{K. pneumoniae}, with initial transmission tree inferred with SeqTrack, an HKY substitution model, and prior infectivity and sampling distributions uniform over the maximum exposure interval observed.
SCOTTI is run for $10^8$ MCMC iterations under an HKY substitution model and with between zero and two non-sampled hosts.
In all cases we checked MCMC convergence with the likelihood trace and effective sample size.
These datasets are provided in \nameref{S1_Data}, and can also be downloaded from https://bitbucket.org/nicofmay/scotti/.
From the same link, SCOTTI source files and executables can be freely downloaded.

\section*{Supporting Information}

\subsection*{S1 Text}
\label{S1_Text}
{\bf Supplementary Text.} The supplementary text contains further details of the methods and analyses, in particular all supplementary figures and tables.

\subsection*{S1 Data}
\label{S1_Data}
{\bf Supplementary Data.} File containing information to replicate results.

\section*{Acknowledgments}
We thank Xavier Didelot, David Eyre, Thibaud Jombart, Erik Volz and Nicole Stoesser for comments and suggestions on early versions of the manuscript.

\nolinenumbers


\newpage

\setcounter{table}{0}
\makeatletter 
\renewcommand{\thetable}{\@Alph\c@table}

\setcounter{figure}{0}

\makeatletter 
\renewcommand{\thefigure}{\@Alph\c@figure}
\makeatother

\section*{S1 Text}

\newpage

\begin{figure}
\caption{{\bf Different representations of transmission events and phylogeny.}
In our manuscript and software, we use several graphical representations to emphasize different aspects of the transmission and evolutionary histories of pathogens
A) To jointly show the evolutionary history of pathogen lineages and transmission events, we adopt ``nested" trees.  Black boxes represent different hosts (here H1, H2, and H3) and their exposure intervals, limited by the top and bottom edges of each box.
Transmission between hosts is represented by blue tubes.
Red dots are sequence samples, and red lines represent the pathogen phylogeny.
B) To focus on the transmission events only, we use ``beanbag" trees. Each host is represented as a circle, with arrows from donor to recipient host.
C) Standard phylogenetic tree relating the sampled sequences.
D) To represent transmission and evolutionary history simultaneously without including epidemiological data, we use ``Maypole" trees. The phylogenetic tree representing the evolutionary history of the pathogen is annotated with colours (one colour for each host) representing the host within which the lineage is inferred to have been. Transition from one colour to another represents transmission between hosts. In this supplement, a similar graphical format is used, where the tree represents the transmission tree enriched with epidemiological data.}
\hspace{-0.cm}\includegraphics[width=0.99\textwidth]{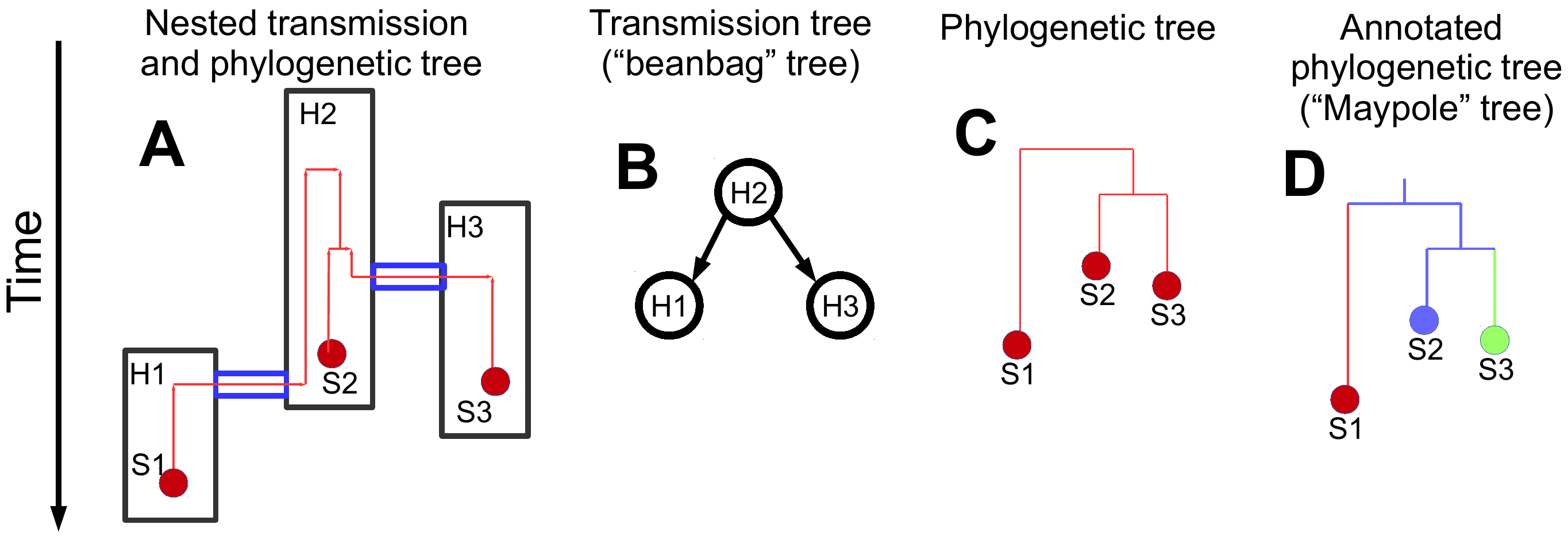}
\label{exampleTrees}
\end{figure}

\begin{figure}
\caption{{\bf Comparison of approaches to transmission inference from genetic data.}
Several methods to reconstruct transmission from genetic and epidemiological data have been proposed in literature, and here we attempt a comparison and summary of their features.
Each row represents a method for inferring transmission history, and each column represents a feature of the model. 
A red ``X" means that the feature is not included, while a green ``V" means that the feature is allowed.
``$-$" methods without an explicit phylogenetic structure can indirectly account for phylogenetic uncertainty.}
\hspace{-4.cm}\includegraphics[width=1.3\textwidth]{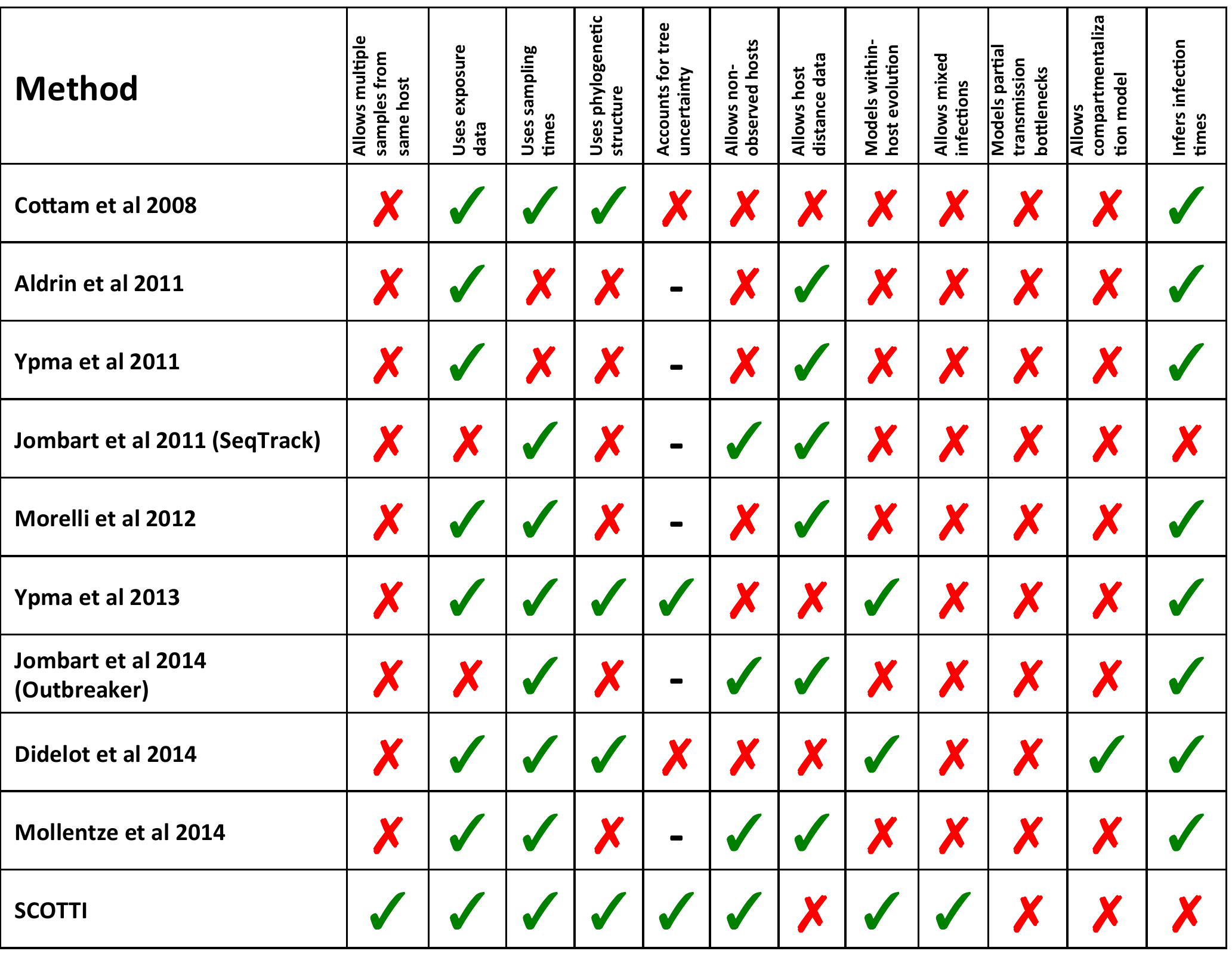}
\label{Figure1}
\end{figure}

\begin{figure}
\caption{{\bf Accuracy of Reconstructed transmissions in the base simulation scenario.}
SCOTTI shows overall higher accuracy than Outbreaker in the base simulation setting.
Coloured trees represent the (fixed) simulated transmission trees, with one colour associated to each host, and internal nodes corresponding to infection events and times, while tips represent infection clearance times.
\textbf{A)} transmission history 1, \textbf{B)} transmission history 2.
In both plots the base simulation setting is considered.
The numbers show the accuracy of the infection origin inference for each sampled host.
Statistics for each origin are plotted below the branch at which top the respective transmission event occurs.
For example, in \textbf{A}, statistics regarding the origin of infection of host M are plotted below the branch representing host M.
Statistics regarding index hosts (K in \textbf{A} and P1 in \textbf{B}) are shown next to the root.
The origin of infection of a host is defined as either the donor host, if it is sampled, or a general non-sampled origin otherwise. 
The non-bracketed numbers represent replicates (out of a total of 100) for which the considered origin has been correctly inferred.
The numbers in brackets are the average posterior support for the corresponding correct origin over all replicates.
For each considered origin, each row represents the results from one of the inference methods (``S1" represents SCOTTI with 1 sample, ``S2" SCOTTI with two samples, and ``O" Outbreaker).
In blue are results for a strong bottleneck (equivalent to the drift of $100 \, N_e$ generations), in red for a weak bottleneck (equivalent to the drift of $N_e$ generations).}
\hspace{-6.92cm}\includegraphics[width=1.61\textwidth]{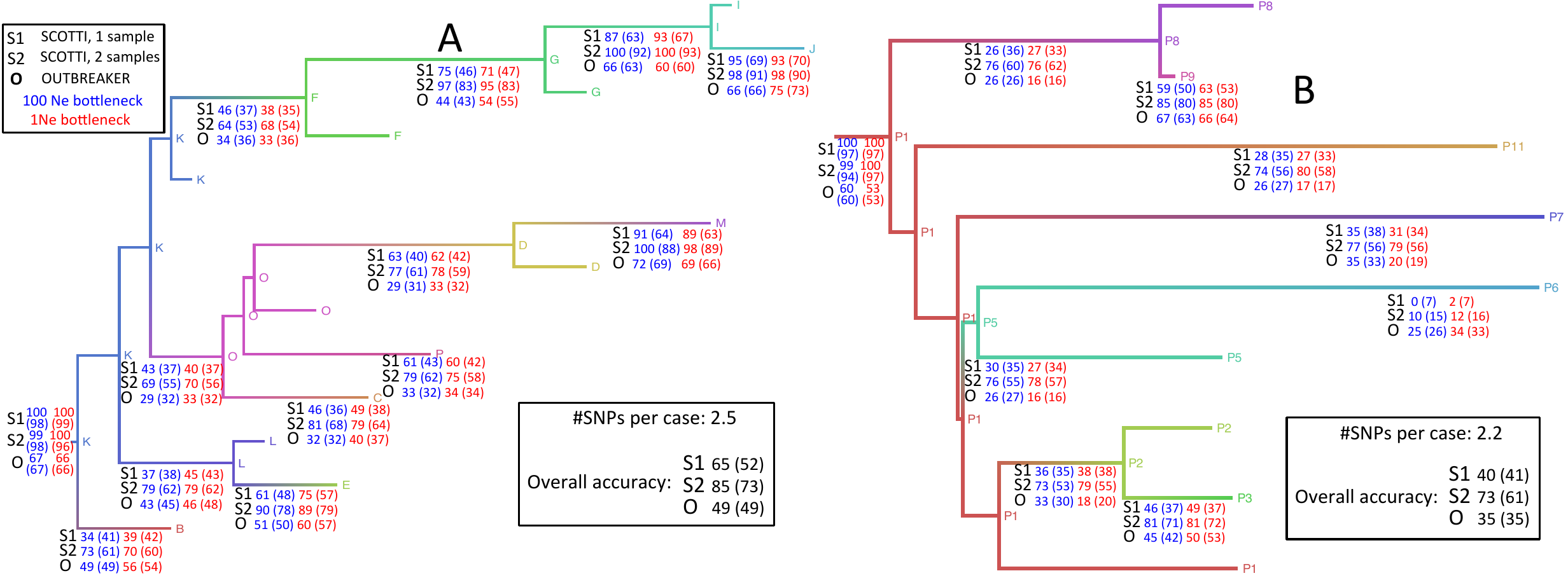}
\label{Base}
\end{figure}

\begin{figure}
\caption{{\bf Summary of errors in transmission inference.}
While error in Outbreaker is mostly attributable to the inference of direct transmission between the wrong pair of sampled hosts, in SCOTTI it is more often due to the incorrect attribution of infection source to non-sampled hosts. 
Pathogen sequence evolution was simulated under transmission history 1, used in \textbf{A} and \textbf{C}, and transmission history 2, used in \textbf{B} and \textbf{D}. 
 In \textbf{A} and \textbf{B} bars represent proportions, expressed as percentages, of incorrect inferences of transmission origin (i.e. donor host) over 100 replicates and all transmission events for each method (differentiated by colour as in legend).
 The proportion of error due to attribution of transmission to non-sampled hosts is shaded with hashes, while the proportion of error due to attribution to sampled hosts is not shaded.
 In \textbf{C} and \textbf{D} bars represent average posterior supports, again expressed as percentages, for the incorrect sources over all patients and replicates.
  On the X axis are different simulation scenarios.}
\hspace{-2.0cm}\includegraphics[width=1.19\textwidth]{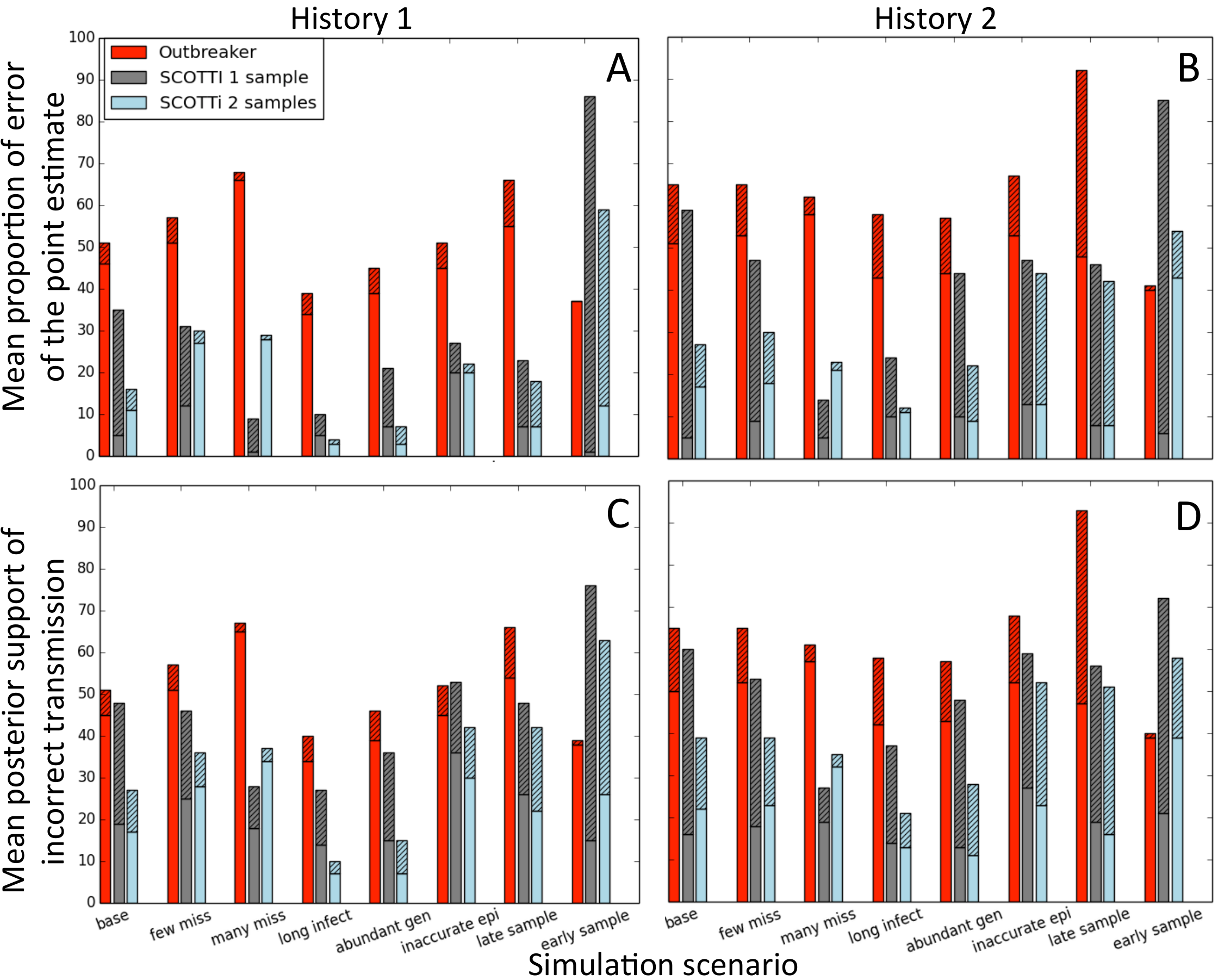}
\label{SummaryError}
\end{figure}

\begin{figure}
\caption{{\bf SCOTTI and Outbreaker accuracy with non-sampled hosts.}
SCOTTI shows overall higher accuracy than Outbreaker in the simulation scenarios with some non-sampled hosts.
Trees represent simulated transmission trees, internal nodes correspond to infection events and times, and tips represent infection clearance times.
The numbers represent inference accuracy as described in Figure~\ref{Base}.
\textbf{A)} Transmission history 1 and one non-sampled host (O).
\textbf{B)} Transmission history 2 and one non-sampled host (P5).
\textbf{C)} Transmission history 1 and three non-sampled hosts (O, G, and H).
\textbf{D)} Transmission history 2 and three non-sampled hosts (P1, P5, and P8).}
\hspace{-6.92cm}\includegraphics[width=1.61\textwidth]{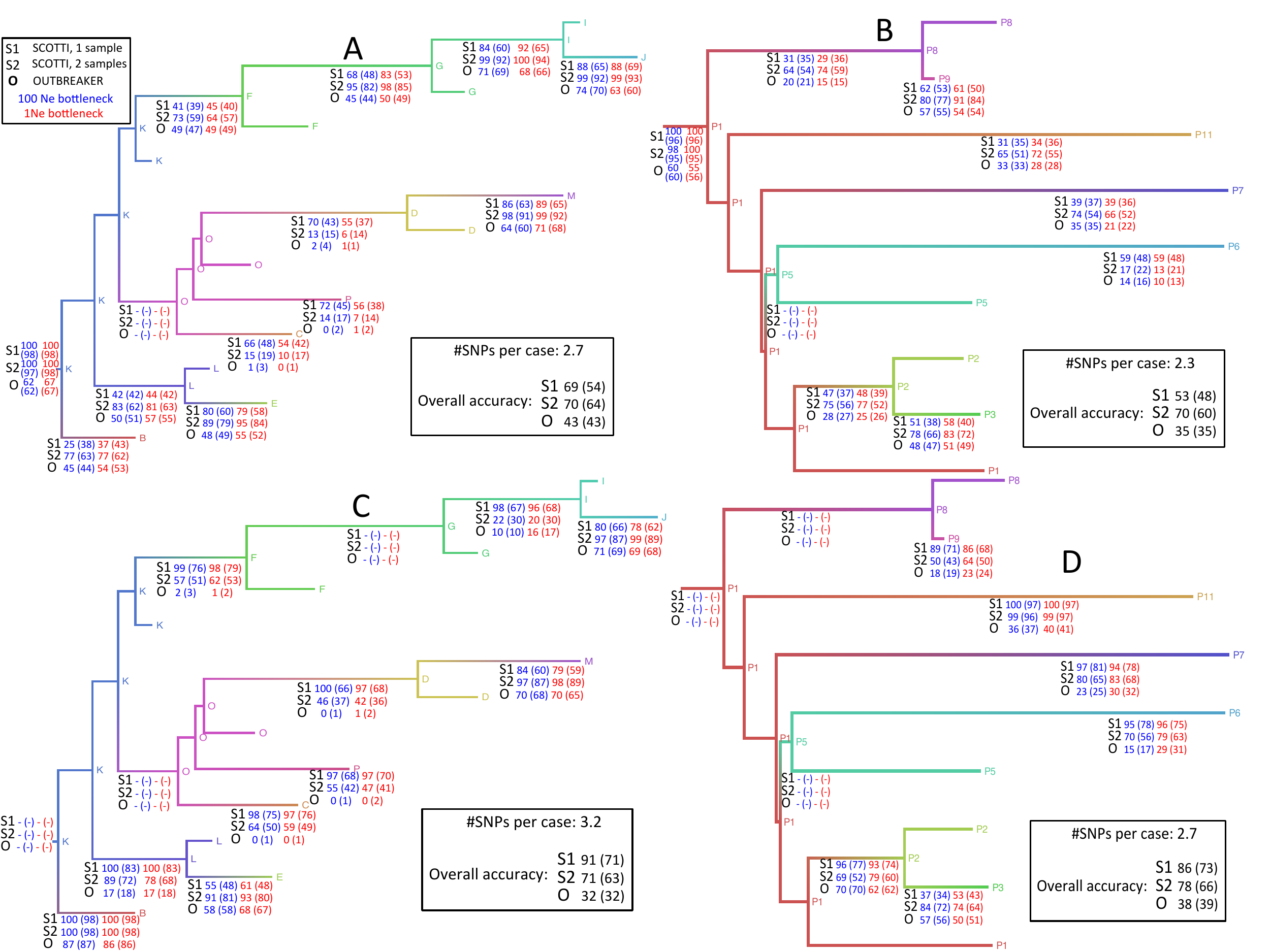}
\label{Missing}
\end{figure}

\begin{figure}
\caption{{\bf Increased accuracy with more genetic variation.}
Both SCOTTI and Outbreaker show increased accuracy when more genetic variation (and so phylogenetic signal) is provided, and SCOTTI shows overall higher accuracy than Outbreaker.
Trees, internal nodes and tips have the same respective meanings as those in Figure~\ref{Missing}.
The numbers represent inference accuracy as described in Figure~\ref{Base}.
\textbf{A)} Transmission history 1 and long infection time (average time of infection $10\, N_e$ generations instead of $2\, N_e$).
\textbf{B)} Transmission history 2 and long infection time.
\textbf{C)} Transmission history 1 and abundant genetic data (15000 base pairs instead of 1500).
\textbf{D)} Transmission history 2 and abundant genetic data.}
\hspace{-6.92cm}\includegraphics[width=1.61\textwidth]{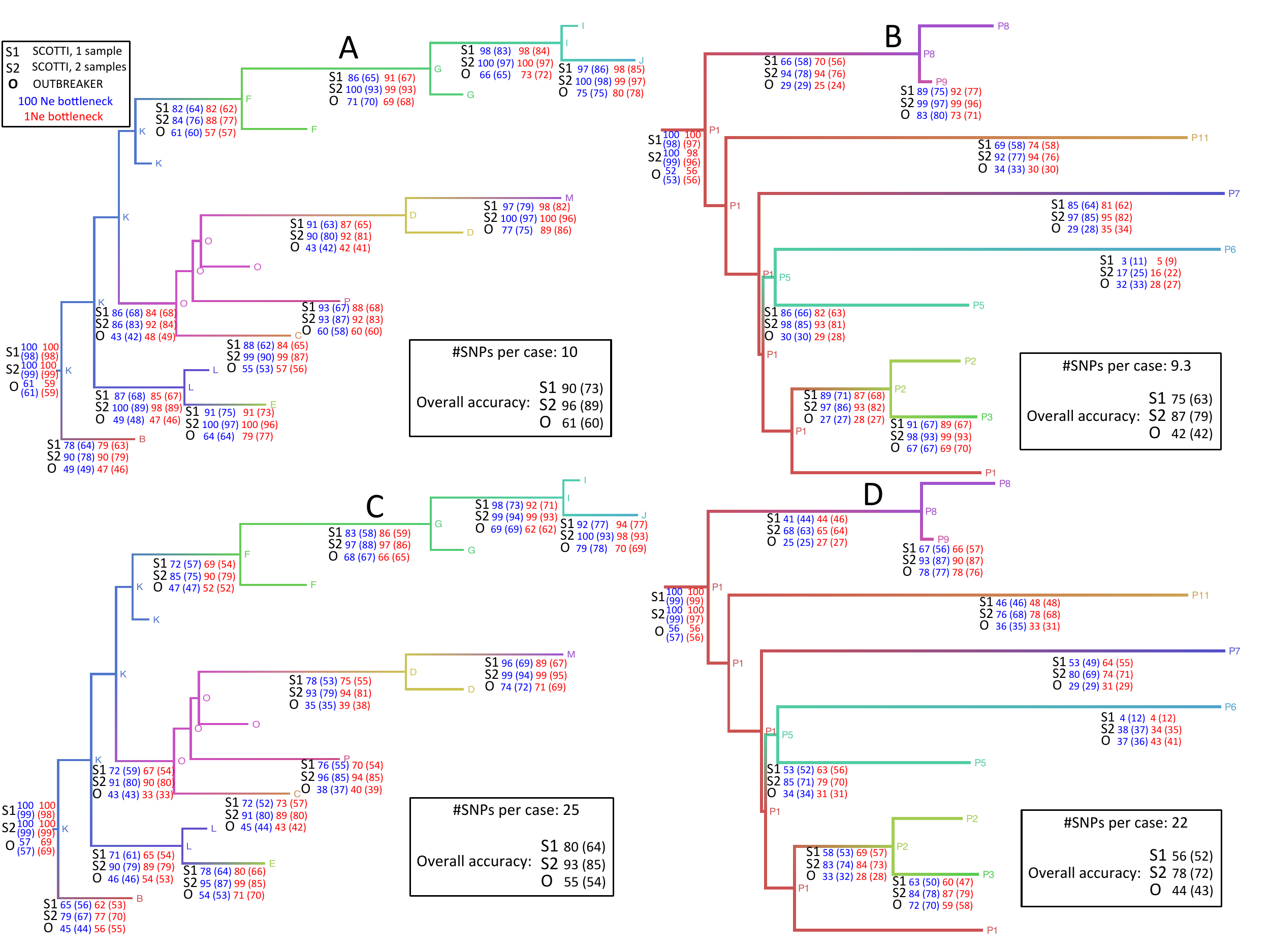}
\label{MoreData}
\end{figure}

\begin{figure}
\caption{{\bf Accuracy of host exposure has limited effect on SCOTTI.}
When provided with inaccurate epidemiological data (exposure intervals are double in length than true ones) SCOTTI is not considerably affected, and still shows higher accuracy than Outbreaker.
Trees, internal nodes and tips have the same respective meanings as those in Figure~\ref{Missing}.
The numbers represent inference accuracy as described in Figure~\ref{Base}.
\textbf{A)} Transmission history 1 and inaccurate introduction and removal times (exposure intervals are double in length than the truth). 
\textbf{B)} Transmission history 2 and inaccurate introduction and removal times.}
\hspace{-6.92cm}\includegraphics[width=1.61\textwidth]{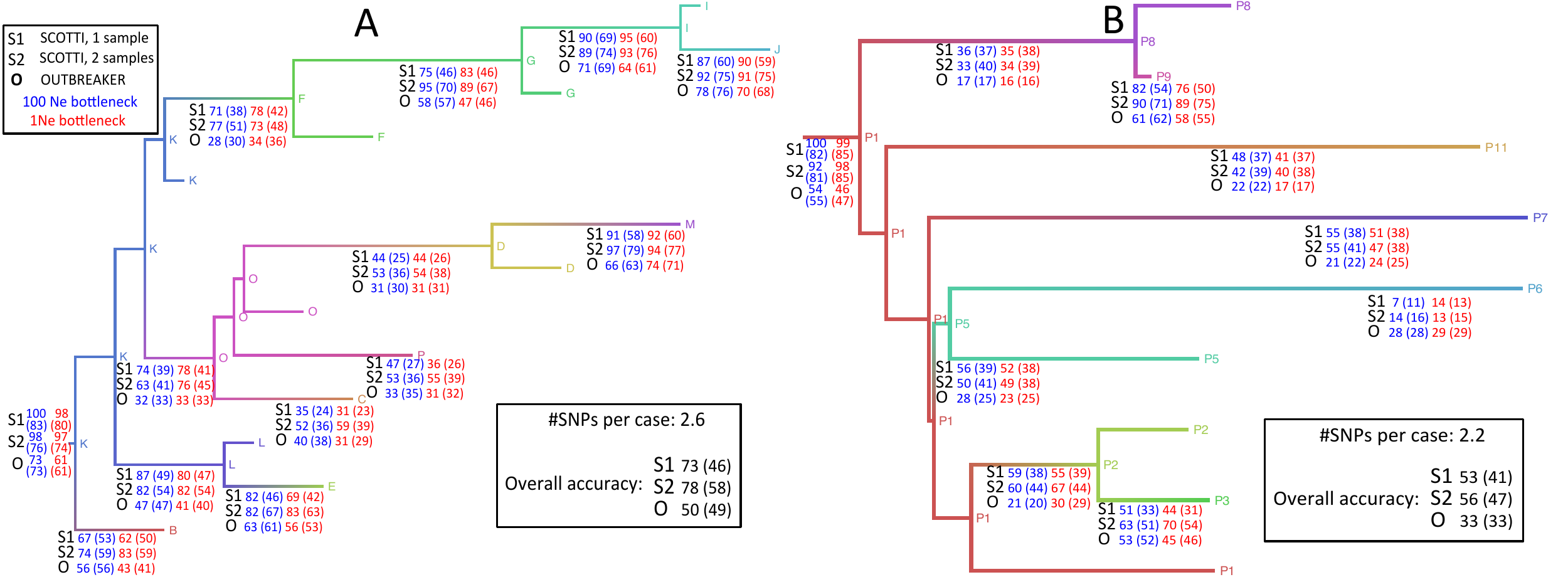}
\label{Inaccurate}
\end{figure}

\begin{figure}
\caption{{\bf Effects of sampling times on the reconstruction of transmission.}
With late sampling times (close to infection clearance) SCOTTI shows higher accuracy than Outbreaker, which has high error rates.
With samples collected early in infection, instead, SCOTTI has a noticeable decrease in accuracy, and becomes less accurate than Outbreaker.
Trees, internal nodes and tips have the same respective meanings as those in Figure~\ref{Missing}.
The numbers represent inference accuracy as described in Figure~\ref{Base}.
\textbf{A)} Transmission history 1 and late sampling (at host clearance).
\textbf{B)} Transmission history 2 and late sampling.
\textbf{C)} Transmission history 1 and early sampling (close to infection time).
\textbf{D)} Transmission history 2 and early sampling.}
\hspace{-6.92cm}\includegraphics[width=1.61\textwidth]{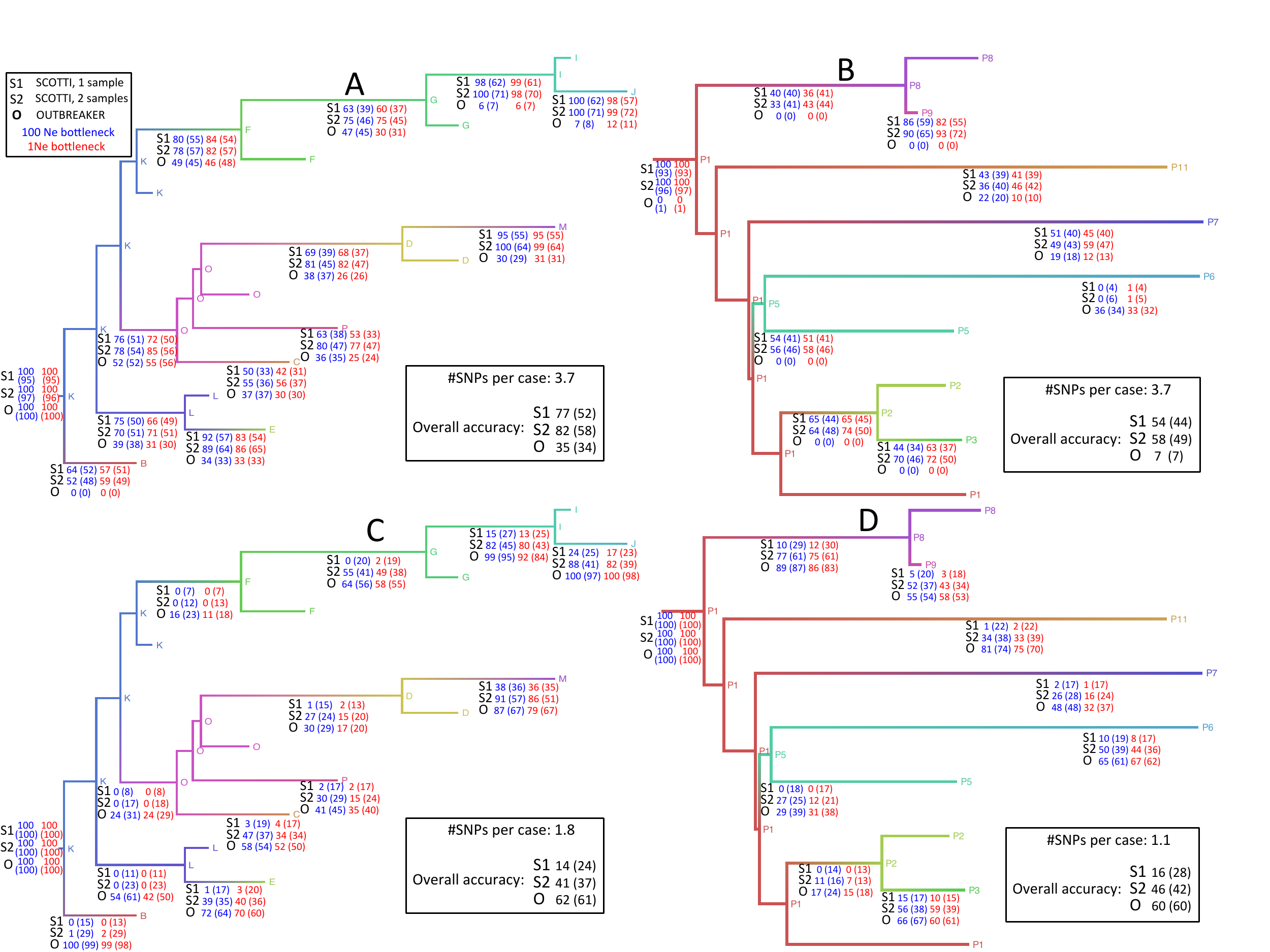}
\label{Sampling}
\end{figure}

\begin{figure}
\caption{{\bf Epidemiological and genetic information from the FMDV outbreak  \cite{COTTETAL08}.}
Details of the FMDV outbreak dataset considered. This figure is reproduced from \cite{COTTETAL08}.
\textbf{A)} Connecting lines represent a nucleotide substitution, thicker lines represent non-synonymous substitutions, with substitutions indicative of adaptation to cell culture coloured green. Sequenced haplotypes (red circles), and putative ancestral virus haplotypes (white circles) are shown. \textbf{B)} Lesion age derived infection profiles of holdings overlaid with the outbreak virus geneology. The orange shading estimates the time when animals with lesions were present from the oldest lesion age at post-mortem. For IP2c, there were no clinical signs of disease. The light blue shading represents incubation periods for each holding, estimated to begin no more than 14 days prior to appearance of lesions. The dark blue shading is the infection date based on the most likely incubation time for this strain of 2-5 days. Each UK 2007 outbreak virus haplotype is plotted according to the time the sample was taken from the affected animal (X axis).}
\hspace{-3.92cm}\includegraphics[width=1.3\textwidth]{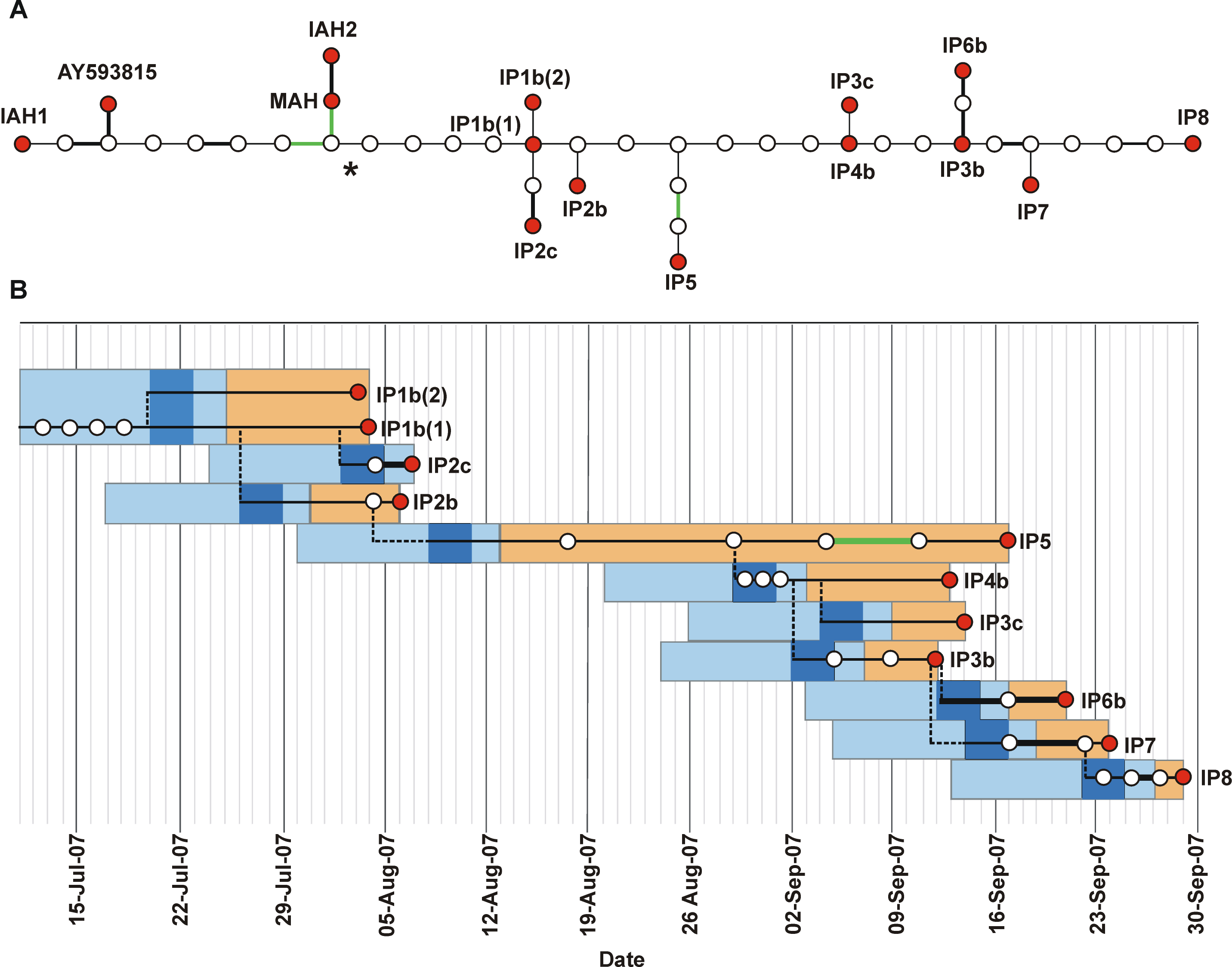}
\label{FMDV_Epi}
\end{figure}

\begin{figure}
\caption{{\bf Presence of patients within the wards affected by the \emph{K. pneumoniae} outbreak \cite{STOEETAL14}.}
Timeline of \emph{K. pneumoniae} patients exposures, including individuals who were both part of epidemiologically defined clusters and had genetically linked outbreak strains. This figure is reproduced from \cite{STOEETAL14}.}
\hspace{-3.92cm}\includegraphics[width=1.2\textwidth]{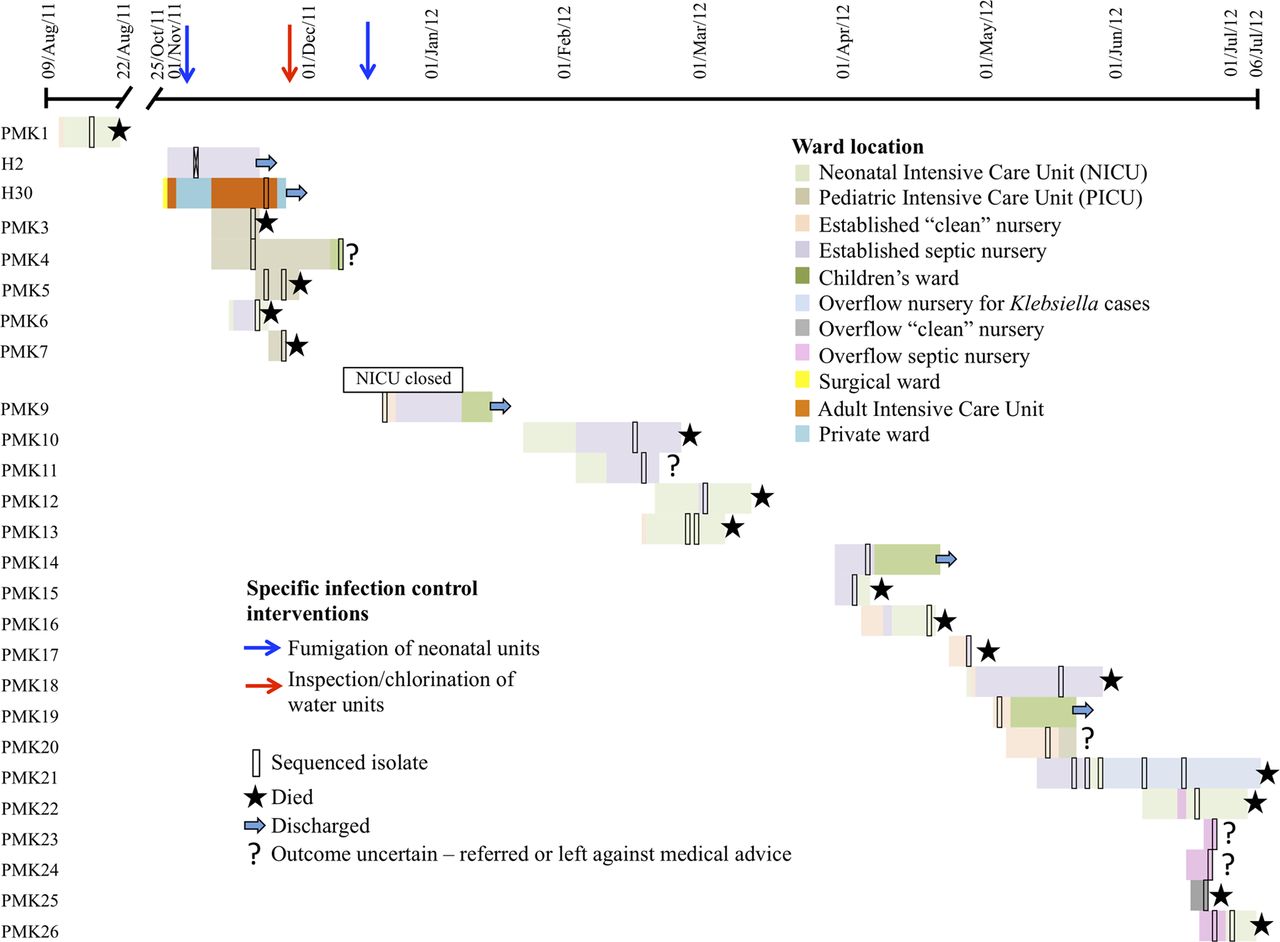}
\label{KlebEpi}
\end{figure}

\begin{table}[h]
\caption{\textbf{Symbols used in model description.}}
\begin{tabular}{ll}
\hline
$D$ & set of all hosts \\
$n_D$ & size of $D$ (total number of hosts) \\
$d_r$ & removal time  \\
$d_i$ & introduction time  \\
{\bf $E$} & set of all exposure time information  \\
$m$ & transmission rate  \\ 
$N_e$ & within-host effective population size  \\ 
$I$ & set of samples  \\ 
$s_i$ & genetic sequence of sample $i$  \\ 
$t_i$ & time of collection of sample $i$  \\ 
$l_i$ & host from which sample $i$ is collected  \\ 
{\bf $\mu$} & substitution rate matrix  \\ 
$T$ & phylogeny relating the samples  \\ 
$M$ & transmission history  \\ 
$A_i=[\alpha_{i-1},\alpha_i]$ & time interval between events  \\ 
$\tau_i=\alpha_i-\alpha_{i-1}$ & length of time between events \\
$\Lambda_i$ & set of extant lineages at interval $i$ \\
$A_{i1}$ and $A_{i2}$ & sub-intervals of $A_i$ \\
$D_i$ & number of hosts exposed at interval $i$ \\\hline
\end{tabular}
\label{Symbols}
\end{table}

\end{document}